\documentclass[lettersize,journal]{IEEEtran}
\usepackage{amsmath,amsfonts,amssymb,mathtools}
\usepackage{algorithmic}
\usepackage{algorithm}
\usepackage{array}
\usepackage{microtype}
\usepackage{subcaption}
\usepackage{balance}
\usepackage[colorlinks]{hyperref}
\hypersetup{citecolor=blue}
\usepackage{textcomp}
\usepackage{stfloats}
\usepackage{url}
\usepackage{verbatim}
\usepackage{graphicx}
\usepackage[nocompress]{cite}
\usepackage{tikz}
\usetikzlibrary{positioning,arrows.meta,calc,fit,backgrounds,shapes.geometric}
\definecolor{gDark}{gray}{0.20}
\definecolor{gMid}{gray}{0.45}
\definecolor{gLight}{gray}{0.80}

\newcommand{\Ci}{\operatorname{Ci}}
\newcommand{\Si}{\operatorname{Si}}
\newcommand{\Herm}{\operatorname{Herm}}
\newcommand{\CN}{\mathcal{CN}}
\newcommand{\E}{\mathbb{E}}
\newcommand{\R}{\mathbb{R}}
\newcommand{\C}{\mathbb{C}}

\usepackage[font=small,skip=2pt]{caption}

\usepackage{enumitem}
\setlist{nosep}

\begin{document}

\title{Electromagnetic-Aware Fluid Antenna Array}

\author{Zhentian Zhang, Yuanhui Wu,
	Hao Jiang,~\IEEEmembership{Senior Member,~IEEE}, \\Zhen Chen,~\IEEEmembership{Senior Member,~IEEE}, Zaichen Zhang,~\IEEEmembership{Senior Member,~IEEE}
	\vspace{-10mm}
	
	\thanks{ }
	
	\thanks{Zhentian Zhang and Zaichen Zhang are with the National Mobile Communications Research Laboratory, Southeast University, Nanjing, 210096, China. (e-mail: zhentianzhangzzt@gmail.com, zczhang@seu.edu.cn).}
	\thanks{Zhen Chen is with the College of Cyber Security, Jinan University, Guangzhou, 510632, P. R. China (e-mail: chenz.scut@gmail.com)}
	\thanks{Hao Jiang is with the School of Cyber Science and Engineering, Southeast University, Nanjing 210096, P. R. China (e-mail: jiang.hao@seu.edu.cn).}
	\thanks{Yuanhui Wu is with the College of Artificial Intelligence, Nanjing University of Information Science and Technology, Nanjing 210044, China (e-mail: 202412621447@nuist.edu.cn)}
	\thanks{Corresponding authors: Zaichen Zhang}
}

\maketitle
\begin{abstract}
	Fluid antenna arrays (FAAs) offer a promising means of exploiting spatial degrees 
	of freedom through adaptive port positioning. However, most existing communication 
	models treat antenna ports as independent channel samples and therefore overlook 
	the electromagnetic coupling that fundamentally governs compact apertures. This 
	paper develops an electromagnetic-aware current-domain framework for planar FAAs. 
	The proposed model integrates position-dependent multiport impedance, mutual 
	coupling, radiated and accepted power, source-voltage feasibility, and channel 
	variations into a unified baseband-compatible description. The framework is 
	implementation-agnostic: the closed-form half-wave-dipole model adopted in this 
	paper is only one instance and can be replaced by full-wave, measured, or surrogate 
	impedance and embedded-pattern models. Building on this framework, we formulate 
	two optimization-oriented design problems. The first addresses single-beam 
	superdirective beamforming through the joint optimization of port currents and 
	positions under sidelobe, current, voltage, and geometry constraints. The second 
	maximizes the multi-user weighted sum rate via current-domain precoding and 
	position optimization under accepted-power, current, voltage, and spacing 
	constraints. In both cases, the electromagnetic model is not applied as an 
	after-design correction, but is incorporated directly into tractable alternating 
	algorithms with convex current or precoding subproblems and reduced-gradient 
	geometry updates. Simulation results demonstrate that, when properly modeled, 
	mutual coupling can be exploited as a valuable design resource, enabling lower 
	sidelobes and persistent sum-rate gains over fixed-grid and random fluid-antenna 
	benchmarks.
\end{abstract}


\section{Introduction}
\subsection{Background and Related Work}
\IEEEPARstart{F}{luid} antenna systems (FAS) \cite{wong2020_fas_limits, fas-twc-21} have emerged as a cornerstone of reconfigurable antenna technologies, reshaping the foundations of future wireless network design \cite{tu1,tu2,tu3}. FAS is a hardware-agnostic system concept that treats the antenna as a reconfigurable physical-layer resource, thereby broadening the system design space and harnessing spatial diversity from either random fading \cite{fas-twc-21} or geometric structure \cite{FAA1}. As a novel theoretical framework, FAS has catalyzed a paradigm shift in next-generation wireless networks, establishing foundational results across multiple access \cite{MA1,MA2}, random access \cite{FBL1,FBL2}, multi-carrier systems \cite{OFDM1,OFDM2}, low-latency communications with finite blocklength \cite{FBL3,FBL4}, beamforming \cite{beam1,FAA2}, and novel signal processing \cite{wt0,zzt_sp1,zzt_sp2,wt1,wt2,wt3}. Early FAS research primarily adopted a fading-domain perspective, exploiting the distinctive plateau and deep-fading characteristics of the channel envelope to extract diversity gains and improve communication reliability. While random fading provides substantial spatial diversity, particularly for interference suppression through the creation of interference nulls among all users, its effective exploitation generally requires relatively accurate channel state information (CSI) in both single-user and multi-user settings. Notably, the CSI can be robustly reconstructed by exploiting the correlation among ports \cite{CSI1} and realized through efficient algorithms \cite{zzt_sp2}.

A complementary and arguably more fundamental dimension of FAS has recently come to the fore: the \emph{geometric diversity} inherent in a fluid antenna array (FAA) \cite{FAA1,FAA2}. In contrast to fading-based benefits, this diversity incurs little overhead to exploit, as it stems from a deliberately reconfigurable geometric structure. The reconfigurable geometry directly governs the effective aperture, spatial sampling structure, and spatial frequency content of the array, and hence its radiation pattern, enabling performance that is not merely quantitatively improved but qualitatively distinct from what fixed apertures permit. As established in \cite{FAA1,FAA2} for both linear and planar FAA topologies, such geometric flexibility serves as the enabling mechanism for fine-grained radiation pattern control, including beamforming gain enhancement, directivity shaping, and, most critically for interference-limited systems, sidelobe suppression.

\subsection{Challenges and Motivations}
Most existing studies model an FAS or FAA as an ideal position-flexible radiator whose ports sample the wireless channel over a continuous aperture, and accordingly attribute the performance limit mainly to the spatial correlation imposed by the propagation environment. While this abstraction is useful for revealing the potential of FASs, it does not fully capture the electromagnetic (EM) behavior \cite{EM0,EM1,EM2} of practical implementations. For instance, from the perspective of EM modeling \cite{Balanis2016}, mutual coupling is the passive re-radiation that occurs when two elements are placed in close proximity: port-A radiates while port-B re-radiates the signal from port-A owing to their close placement, an effect that is widely present yet commonly overlooked in state-of-the-art FAS and FAA studies. Although mutual coupling is traditionally regarded as detrimental \cite{Balanis2016,gupta1983effect}, the reality proves to be otherwise, as the outcome depends critically on how it is modeled and harnessed \cite{Damico2024,Ramirez2026BeamspaceFAS,Ramirez2025MetasurfaceFAS}.

Recent EM-compliant studies have further revealed that practical FASs should be modeled beyond the ideal spatial-sampling abstraction. For metasurface-based FASs, the antenna can be represented as a multiport circuit network in which load tuning, impedance matching, radiation efficiency, reflections, and inter-element mutual coupling are embedded into the equivalent baseband channel and its covariance structure \cite{Ramirez2025MetasurfaceFAS}. In this beamspace view, each FAS port corresponds to a feasible antenna configuration rather than a purely geometric position, and a properly designed configuration codebook can reshape the effective channel response and even project the received signal away from dominant interference subspaces \cite{Ramirez2026BeamspaceFAS}. A similar but more physically explicit interpretation arises in reconfigurable pixel antennas (RPAs) \cite{Wong2026RPAFAS}, where the ON/OFF states of inter-pixel switches modify the current flow over the pixelated aperture. The resulting mutual coupling among pixels is no longer merely a hardware impairment. Instead, it acts as an intrinsic analog electromagnetic signal-processing mechanism that mixes the incident fields through configuration-dependent surface currents, thereby synthesizing diverse radiation patterns and controllable channel correlations. {\em These observations suggest that the performance headroom of practical FASs is closely tied to how accurately the EM coupling network is modeled and how effectively it is exploited.} In particular, by jointly designing antenna configurations, matching constraints, and communication objectives, dense reconfigurable apertures can leverage mutual coupling to generate richer effective channels, improve interference avoidance, and unlock gains that remain invisible under conventional position-only FAS models.

These recent results indicate that electromagnetic modeling is not merely a tool for hardware verification, but a key enabler for uncovering the remaining performance headroom of FASs. If mutual coupling is accurately modeled and judiciously utilized, considerable gains can still be extracted from practical fluid-antenna architectures. Nevertheless, all existing works investigate mutual coupling within specific circuit implementations, such as metasurfaces \cite{Ramirez2025MetasurfaceFAS} or pixel antennas \cite{Wong2026RPAFAS}. Establishing a more universally applicable and practically feasible electromagnetic model for FAS therefore remains a critical open problem. Although fluid antenna arrays provide a new spatial degree of freedom through port-position adaptation, their practical performance is fundamentally shaped by electromagnetic effects such as mutual coupling, radiation loss, source mismatch, and voltage feasibility. The main challenge is therefore to retain these physical effects while still arriving at communication design problems that can be solved efficiently.

\subsection{Contributions}
The main contributions of this paper are summarized as follows.

\begin{enumerate}
	\item \textbf{General EM-aware FAA model:}
	We develop, to the best of our knowledge, the first general current-domain framework that converts a position-dependent multiport electromagnetic model of a planar FAA into a communication-oriented optimization model. The proposed formulation jointly captures mutual coupling, radiated and accepted power, source-voltage feasibility, current robustness, geometry constraints, and position-dependent channels. The model is not tied to a specific antenna approximation: the analytical dipole impedance adopted in this paper can be replaced by full-wave, measurement-based, or surrogate electromagnetic models. This establishes a physically consistent bridge between antenna modeling and system-level optimization.
	
	\item \textbf{Optimization-ready single-beam beamforming:}
	Based on the proposed model, we formulate an EM-aware single-beam superdirective beamforming problem with jointly optimized port currents and antenna positions. The formulation recasts the physical beam design as a constrained optimization problem subject to sidelobe, current, source-voltage, and spacing constraints. For fixed antenna positions, the current design problem is convex, while the geometry is updated through reduced-gradient sensitivity analysis and a spacing-preserving line search. The results show that mutual coupling and physical constraints not only limit the feasible set, but also regularize the excitation design and help achieve lower sidelobes than baseline schemes.
	
	\item \textbf{Optimization-ready multi-user sum-rate maximization:}
	We further extend the framework to downlink multi-user transmission by formulating an EM-aware weighted sum-rate maximization problem. The current-domain precoder and the configurable-port geometry are jointly designed under accepted-power, current, source-voltage, and spacing constraints. For fixed positions, the nonconvex precoding problem is handled via fractional programming, while the antenna positions are updated using reduced-gradient information that incorporates both channel and electromagnetic derivatives. Simulation results show that optimized FAA geometries provide persistent sum-rate gains over fixed-grid and random geometries, confirming that properly modeled mutual coupling can be exploited to reshape effective channels and improve multi-user interference management.
\end{enumerate}

The remainder of this paper is organized as follows. Section~\ref{sec:em_fas} develops the EM-aware FAA model. Section~\ref{sec:single_beam} presents the single-beam superdirective beamforming design. Section~\ref{sec:multiuser} develops the multi-user weighted sum-rate maximization framework. The simulation section evaluates the proposed designs and compares them with fixed-grid and random FAA benchmarks, followed by the conclusion.

{\em Notation:} Bold lowercase and uppercase letters denote vectors and matrices, respectively. The superscripts $(\cdot)^{\mathrm T}$ and $(\cdot)^{\mathrm H}$ denote transpose and conjugate transpose. The operators $\|\cdot\|_2$, $\mathbb{E}[\cdot]$, $\operatorname{Re}\{\cdot\}$, and $\operatorname{Herm}(\cdot)$ denote the Euclidean norm, expectation, real part, and Hermitian part, respectively. The imaginary unit is denoted by $j$, and $\mathbf{I}_N$ denotes the identity matrix of size $N$.

\begin{figure*}[t]
	\centering
	\resizebox{1.8\columnwidth}{!}{%
		\begin{tikzpicture}[
			font=\sffamily\small,
			>=Stealth,
			aperture/.style ={draw=black, line width=1.0pt},
			gridline/.style ={draw=black, line width=0.3pt, dotted},
			arm/.style      ={draw=black, line width=1.6pt},
			coupling/.style ={draw=black, line width=0.6pt, dash dot dot},
			ring/.style     ={draw=black, line width=0.5pt, densely dashed},
			config/.style   ={->, draw=black, line width=0.7pt},
			farfield/.style ={->, draw=black, line width=1.2pt},
			lead/.style     ={->, draw=black, line width=0.4pt},
			callout/.style  ={font=\sffamily\scriptsize, text=black, align=left},
			portlbl/.style  ={font=\sffamily\scriptsize, inner sep=1pt,
				fill=white, draw=none}
			]
			
			\draw[aperture] (0,0) rectangle (7,5);
			\node[anchor=north west, font=\sffamily\small\itshape] at (0.1,4.95) {Aperture $\mathcal{A}$};
			\node[anchor=south west, font=\sffamily\scriptsize] at (0.1,0.05)
			{planar region where ports may be configured};
			\foreach \x in {1,2,...,6} \draw[gridline] (\x,0)--(\x,5);
			\foreach \y in {1,2,3,4}   \draw[gridline] (0,\y)--(7,\y);
			
			\newcommand{\dipole}[5]{%
				\draw[ring] (#1,#2) circle (0.55);
				\draw[arm] (#1,#2+0.06) -- (#1,#2+0.42);
				\draw[arm] (#1,#2-0.06) -- (#1,#2-0.42);
				\fill[white] (#1,#2) circle (2.4pt);
				\draw[draw=black, line width=0.7pt] (#1,#2) circle (2.4pt);
				\node[portlbl] at ($(#1,#2)+(#4:#5)$) {#3};
			}
			\dipole{1.4}{1.2}{$n_1$}{200}{0.85}   
			\dipole{3.0}{3.4}{$n_2$}{160}{0.85}   
			\dipole{5.2}{1.5}{$n_3$}{ 20}{0.85}   
			\dipole{5.6}{3.6}{$n_4$}{ 30}{0.85}   
			\dipole{2.6}{1.0}{$n_5$}{70}{0.85}    

			\draw[coupling] (1.4,1.2) -- (3.0,3.4);
			\draw[coupling] (3.0,3.4) -- (5.6,3.6);
			\draw[coupling] (1.4,1.2) -- (2.6,1.0);
			\draw[coupling] (5.2,1.5) -- (5.6,3.6);
			
			\draw[config] (3.0,3.4)++(0.55,0) -- ++(0.6,0);
			\draw[config] (3.0,3.4)++(-0.55,0) -- ++(-0.6,0);
			\draw[config] (3.0,3.4)++(0,0.55) -- ++(0,0.5);
			\draw[config] (3.0,3.4)++(0,-0.55) -- ++(0,-0.5);
			
			\draw[farfield] (7,2.5) -- (8.4,3.4);
			\draw[farfield] (7,2.5) -- (8.5,2.5);
			\draw[farfield] (7,2.5) -- (8.4,1.6);
			\draw[draw=black, line width=0.4pt, densely dashed] (7.7,2.5) ++(0:0) arc (28:-28:1.1);
			
			\node[callout, anchor=west] (Ldip) at (8.4,4.6)
			{\textbf{Half-wave dipole}\\(\,$\lambda/2$ thin-wire\\element / port $n$\,)};
			\draw[lead] (Ldip.west) to[bend right=15] (3.0,3.82);
			
			\node[callout, anchor=west] (Lfeed) at (8.7,1.0)
			{\textbf{Feed port}\\center-fed terminal\\(current $I_n$, voltage $v_n$)};
			\draw[lead] (Lfeed.west) to[bend left=10] (5.2,1.5);
			
			\node[callout, anchor=east] (Lmut) at (-0.4,3.6)
			{\textbf{Mutual coupling}\\$Z_{mn}=Z(\lVert p_m-p_n\rVert)$\\distance-only EM model};
			\draw[lead] (Lmut.east) to[bend left=10] (2.2,2.3);
			
			\node[callout, anchor=east] (Lmin) at (-0.4,1.6)
			{\textbf{Min spacing} $d_{\min}$\\exclusion ring\\(avoids overlap /\\super-directivity)};
			\draw[lead] (Lmin.east) to[bend right=10] (0.95,1.05);
			
			\node[callout, anchor=south] (Lmov) at (3.0,5.6)
			{\textbf{Configurable position} $p_n=(x_n,y_n)$ \;---\; differentiable over $\mathcal{A}$};
			\draw[lead] (Lmov.south) to[bend left=8] (3.0,4.5);
			
			\node[callout, anchor=north east, align=center,
			draw=black, line width=0.6pt, inner sep=4pt] (LZ) at (-0.4,5.6)
			{\textbf{Multiport EM model}\\
				$\mathbf{Z}(\mathbf{p})\in\mathbb{C}^{N\times N}$\\
				{\scriptsize $\mathrm{Re}\{\mathbf{Z}\}\!\to$ radiated power}\\
				{\scriptsize $\mathrm{Im}\{\mathbf{Z}\}\!\to$ reactive coupling}};
			
			\node[callout, anchor=west] (Lff) at (8.7,2.5)
			{\textbf{Far-field radiation}\\array response\\$\mathbf{a}(\theta,\phi;\mathbf{p})$};
			
		\end{tikzpicture}
	}%
	\caption{Electromagnetic-aware structure of a planar FAA. Half-wave dipole elements (ports) can be continuously configured over the planar aperture~$\mathcal{A}$. Their positions~$\mathbf{p}=(x_n,y_n)$ determine the distance-dependent mutual coupling and hence the multiport impedance matrix~$\mathbf{Z}(\mathbf{p})$, whose real part governs radiated power and imaginary part the reactive coupling. Dashed rings denote the minimum-spacing constraint~$d_{\min}$, and the right-hand arrows indicate the position-dependent far-field array response~$\mathbf{a}(\theta,\phi;\mathbf{p})$.}
	\label{fig:faa_em_structure}
	\vspace{-6mm}
\end{figure*}
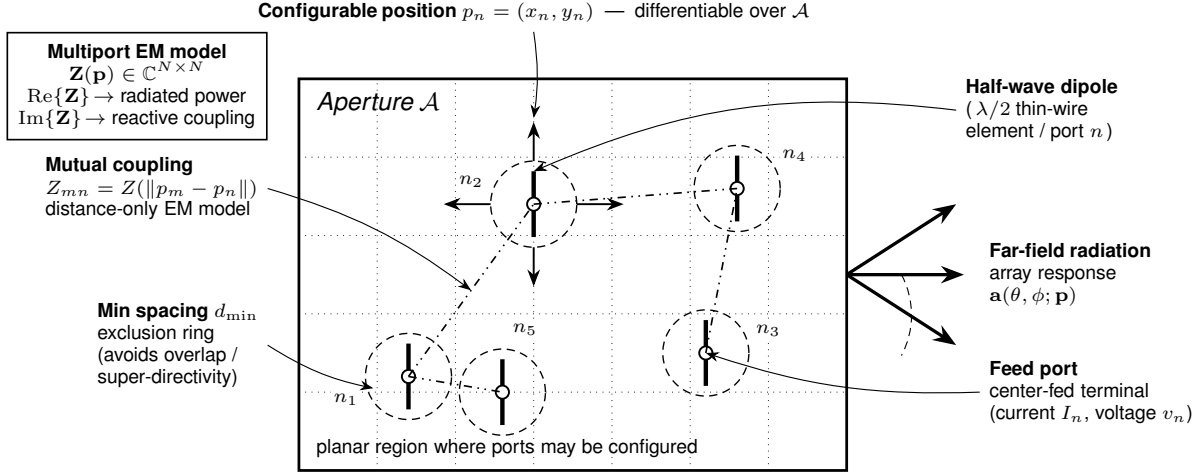
\section{Electromagnetic-Aware FAA Model}
\label{sec:em_fas}

\subsection{Planar FAA Geometry and Modeling Assumptions}

Consider a planar array with a rectangular aperture on the plane $z=0$,
\begin{equation}
	\mathcal{A}=[0,W_x]\times[0,W_y],
\end{equation}
where the planar coordinate and the corresponding three-dimensional center
position of port $n=1,\ldots,N$ are
\begin{equation}
	\mathbf{p}_n=[x_n,y_n]^\mathrm{T}\in\R^2,~
	\mathbf{r}_n=[x_n,y_n,0]^\mathrm{T}.
\end{equation}
The stacked position vector is defined as
\begin{equation}
	\mathbf{p}
	=
	[\mathbf{p}_1^\mathrm{T},\ldots,\mathbf{p}_N^\mathrm{T}]^\mathrm{T}\in\R^{2N},
\end{equation}
whose feasible set is
\begin{equation}
	\label{eq:position_set}
	\mathcal{P}
	=
	\left\{
	\mathbf{p}:
	\begin{array}{l}
		\mathbf{p}_n\in\mathcal{A},\quad n=1,\ldots,N,\\[1mm]
		\|\mathbf{p}_m-\mathbf{p}_n\|_2\ge d_{\min},
		\quad m\neq n
	\end{array}
	\right\}.
\end{equation}

Unless otherwise stated, all phasors are root-mean-square (RMS) quantities.
Specifically, for a real sinusoidal scalar quantity
\begin{equation}
	x(t)=X_{\mathrm{pk}}\cos(\omega t+\varphi),
\end{equation}
the corresponding RMS phasor is defined as
\begin{equation}
	X=\frac{X_{\mathrm{pk}}}{\sqrt{2}}e^{j\varphi},
\end{equation}
or, equivalently,
\begin{equation}
	x(t)=\sqrt{2}\,\Re\left\{X e^{j\omega t}\right\},
	\label{eq:rms_phasor_definition}
\end{equation}
where the time convention $e^{j\omega t}$ is adopted throughout the paper.
Under this RMS convention, the time-averaged power associated with voltage and
current phasors can be expressed without the additional factor $1/2$ that
would otherwise appear when peak phasors are used \cite{pozar2011microwave}.

The analytical impedance model developed below assumes identical, parallel,
center-fed, electrically thin dipoles in free space \cite{Balanis2016}. The
dipoles are oriented perpendicular to the aperture plane, so their translations
over $\mathcal{A}$ produce a side-by-side configuration. Under these
assumptions, the self-impedance is identical across all ports, whereas the
mutual impedance between ports $m$ and $n$ depends only on their
center-to-center distance,
\begin{equation}
	\rho_{mn}=\|\mathbf{p}_m-\mathbf{p}_n\|_2,\quad m\neq n.
\end{equation}
Accordingly, the off-diagonal impedance entries can be written as
\begin{equation}
	Z_{mn}=z_{\mathrm{mut}}(\rho_{mn}),\quad m\neq n.
\end{equation}
This distance-dependent impedance model provides a tractable
electromagnetic-aware approximation for studying the interplay between mutual
coupling and port-position optimization. Nevertheless, mutual coupling can
substantially affect array behavior and should be modeled consistently in
compact or adaptive arrays \cite{gupta1983effect}. For antennas placed near
dielectric bodies, finite ground planes, chassis edges, or other scatterers,
the impedance matrix generally depends on the full electromagnetic environment
and should accordingly be replaced by a full-wave, measured, or surrogate
impedance model \cite{harrington1993field}. The EM-aware structure
of a planar FAA is illustrated in Fig.~\ref{fig:faa_em_structure}.

\subsection{Continuous Impedance Operator and Multiport Network}

Let $\ell_n$ denote the wire contour of antenna $n$, $I_n$ its feed-current
phasor, and $\mathbf{J}_n(\mathbf{s})$ the corresponding current distribution.
Following the reaction-based or induced-EMF formulation, the impedance entry
between ports $m$ and $n$ can be expressed abstractly as
\begin{align}
	\label{eq:impedance_operator}
	Z_{mn}
	&=
	-\frac{1}{I_m I_n}
	\int_{\ell_m}
	\mathbf{J}_m(\mathbf{s})
	\cdot
	\mathbf{E}_n(\mathbf{s})
	\,d\ell,
	\\
	\mathbf{E}_n(\mathbf{s})
	&=
	\int_{\ell_n}
	\mathcal{G}_{E}(\mathbf{s},\mathbf{s}')
	\mathbf{J}_n(\mathbf{s}')
	\,d\ell',
\end{align}
where $\mathcal{G}_{E}$ denotes the current-to-electric-field Green operator,
incorporating both vector- and scalar-potential contributions consistent with
the adopted $e^{j\omega t}$ time convention. This continuous-operator
expression is consistent with standard reaction-based antenna impedance
modeling and method-of-moments formulations \cite{harrington1993field,Balanis2016}.
The resulting multiport relation is
\begin{equation}
	\label{eq:multiport}
	\mathbf{v}=\mathbf{Z}(\mathbf{p})\mathbf{i},
\end{equation}
where $\mathbf{i}=[i_1,\ldots,i_N]^\mathrm{T}$ is the vector of antenna-port
current phasors and $\mathbf{v}$ is the vector of antenna terminal-voltage
phasors.

For reciprocal antennas embedded in a reciprocal electromagnetic environment,
the impedance matrix satisfies $\mathbf{Z}^{\mathrm{T}}=\mathbf{Z}$, that is,
$\mathbf{Z}$ is complex symmetric rather than Hermitian
\cite{harrington1993field,pozar2011microwave}. We decompose the impedance
matrix as
\begin{equation}
	\label{eq:Z_decomposition}
	\mathbf{Z}
	=
	\mathbf{Z}_{\mathrm{em}}
	+
	\mathbf{R}_{\mathrm{loss}},
\end{equation}
where $\mathbf{Z}_{\mathrm{em}}$ collects the radiation and stored-field
contributions, and $\mathbf{R}_{\mathrm{loss}}\succeq\mathbf{0}$ models
conductor, dielectric, and other dissipative losses. For identical antennas
with independent lumped port losses, a common diagonal approximation is
\begin{equation}
	\mathbf{R}_{\mathrm{loss}}=R_{\ell}\mathbf{I}_N,
\end{equation}
where $R_{\ell}\ge 0$ is the equivalent loss resistance at each antenna port.
When ohmic and dielectric losses are neglected, one may set $R_{\ell}=0$.
For a common real reference impedance $Z_{\mathrm{ref}}>0$, the corresponding
scattering matrix is
\begin{equation}
	\mathbf{S}
	=
	\left(\mathbf{Z}-Z_{\mathrm{ref}}\mathbf{I}_N\right)
	\left(\mathbf{Z}+Z_{\mathrm{ref}}\mathbf{I}_N\right)^{-1},
	\label{eq:Z_to_S}
\end{equation}
and the admittance matrix is $\mathbf{Y}=\mathbf{Z}^{-1}$ whenever $\mathbf{Z}$
is nonsingular. The impedance, admittance, and scattering descriptions
constitute standard equivalent representations of the multiport network
\cite{pozar2011microwave}.

\subsection{Mutual Impedance Model for Side-by-Side Half-Wave Dipoles}

For two identical, parallel, side-by-side thin-wire dipoles with center
separation
\begin{equation}
	d_{mn}=\|\mathbf{p}_m-\mathbf{p}_n\|_2,
\end{equation}
the sinusoidal-current induced-EMF approximation yields, for $m\neq n$,
\begin{align}
	\label{eq:mutual_impedance}
	Z_{mn}(d_{mn})
	&=
	\frac{\eta_0}{4\pi}
	\left[
	2\Ci(u_0)
	-\Ci(u_+)
	-\Ci(u_-)
	\right]
	\nonumber\\
	&\quad
	-j\frac{\eta_0}{4\pi}
	\left[
	2\Si(u_0)
	-\Si(u_+)
	-\Si(u_-)
	\right],
\end{align}
where
\begin{subequations}
	\begin{align}
		u_0&=\beta d_{mn},\\
		u_{\pm}
		&=
		\beta
		\left(
		\sqrt{d_{mn}^{2}+L_{\mathrm d}^{2}}
		\pm L_{\mathrm d}
		\right),\\
		\beta&=\frac{2\pi}{\lambda},~
		L_{\mathrm d}=\frac{\lambda}{2}.
	\end{align}
\end{subequations}
Here, $L_{\mathrm d}$ denotes the \emph{total dipole length} rather than the
length of a single arm, and $\eta_0$ is the free-space wave impedance.

Equation~\eqref{eq:mutual_impedance} provides an approximate closed-form
expression for the specified side-by-side geometry. The diagonal entries should
be obtained from the self-impedance model of the actual element, a full-wave
simulation, or a measurement:
\begin{equation}
	Z_{nn}=Z_{\mathrm{self}}+R_{\ell}.
\end{equation}
For an ideal infinitesimally thin dipole of total length $\lambda/2$, the
familiar self-impedance is approximately $73.1+j42.5~\Omega$, whereas a
resonant physical dipole is typically shortened so that its input reactance
is close to zero \cite{Balanis2016,Damico2024}. Under the homogeneous
free-space translation model, $Z_{nn}$ is independent of $\mathbf{p}_n$.

\subsection{Current-Domain Source and Power Formulation}

Each port is driven by an identical Thevenin source impedance
\begin{equation}
	Z_s=R_s+jX_s,~R_s>0.
\end{equation}
Let $\mathbf{v}_s$ denote the vector of open-circuit source-voltage phasors.
By Kirchhoff's voltage law, the source and antenna-port variables satisfy
\begin{equation}
	\label{eq:source_model}
	\mathbf{v}_s
	=
	\left(
	\mathbf{Z}+Z_s\mathbf{I}_N
	\right)\mathbf{i}
	\triangleq
	\mathbf{C}(\mathbf{p})\mathbf{i}.
\end{equation}
This is the standard Thevenin-equivalent representation of a driven multiport
network \cite{pozar2011microwave}. The corresponding source-voltage quadratic
form is
\begin{equation}
	\label{eq:Qv}
	\|\mathbf{v}_s\|_2^2
	=
	\mathbf{i}^\mathrm{H}\mathbf{Q}_v(\mathbf{p})\mathbf{i},~
	\mathbf{Q}_v(\mathbf{p})
	=
	\mathbf{C}^\mathrm{H}(\mathbf{p})\mathbf{C}(\mathbf{p}).
\end{equation}

Define the Hermitian-part operator as
\begin{equation}
	\Herm(\mathbf{A})
	=
	\frac{\mathbf{A}+\mathbf{A}^\mathrm{H}}{2}.
\end{equation}
The radiation-resistance, loss-resistance, and accepted-power matrices are
defined as
\begin{subequations}
	\label{eq:power_matrices}
	\begin{align}
		\mathbf{R}_{\mathrm{rad}}
		&=
		\Herm(\mathbf{Z}_{\mathrm{em}}),\\
		\mathbf{R}_{\mathrm{acc}}
		&=
		\Herm(\mathbf{Z})
		=
		\mathbf{R}_{\mathrm{rad}}
		+
		\mathbf{R}_{\mathrm{loss}}.
	\end{align}
\end{subequations}
These definitions follow the standard network and antenna-power interpretation
of the real part of the impedance matrix \cite{pozar2011microwave,Balanis2016}.
Using RMS phasors, the radiated power, dissipated antenna power, and accepted
antenna power are given by
\begin{subequations}
	\label{eq:physical_powers}
	\begin{align}
		P_{\mathrm{rad}}
		&=
		\mathbf{i}^\mathrm{H}
		\mathbf{R}_{\mathrm{rad}}
		\mathbf{i},\\
		P_{\mathrm{loss}}
		&=
		\mathbf{i}^\mathrm{H}
		\mathbf{R}_{\mathrm{loss}}
		\mathbf{i},\\
		P_{\mathrm{acc}}
		&=
		\mathbf{i}^\mathrm{H}
		\mathbf{R}_{\mathrm{acc}}
		\mathbf{i}.
	\end{align}
\end{subequations}
The power dissipated in the identical source resistances is
\begin{equation}
	P_s=R_s\|\mathbf{i}\|_2^2,
\end{equation}
where the absence of an additional factor $1/2$ follows from the RMS phasor
convention \cite{pozar2011microwave}. Consequently, when $R_s$ is fixed, a
current-norm constraint simultaneously bounds the source-resistance
dissipation.

The current-domain formulation is essential for avoiding a purely algebraic
interpretation of mutual coupling. In closely spaced arrays, mutual coupling
alters the relations among source voltages, port currents, accepted power,
and radiated fields, and must therefore be retained in the physical
optimization model \cite{gupta1983effect,Damico2024}. If the source voltage
$\mathbf{x}$ were instead chosen as the optimization variable with
$\mathbf{i}=\mathbf{C}^{-1}\mathbf{x}$, then, in the absence of a
source-voltage, available-power, matching, or amplifier constraint, the
change of variables $\mathbf{q}=\mathbf{C}^{-1}\mathbf{x}$ would eliminate
$\mathbf{C}^{-1}$ from the optimization problem, thereby concealing the
coupling effect. In this work, the port currents are optimized directly,
while the physically meaningful constraint
\begin{equation}
	\label{eq:voltage_constraint_general}
	\mathbf{i}^\mathrm{H}\mathbf{Q}_v(\mathbf{p})\mathbf{i}
	\le V_{\max}^2
\end{equation}
preserves the influence of the complete complex impedance matrix on source
feasibility.

\subsection{Array Radiation Response and Position-Dependent Channels}

Let $\Omega=(\theta,\phi)$ denote a far-field direction. We define the
current-to-radiation response vector
$\mathbf{b}(\Omega;\mathbf{p})\in\C^{N}$ such that the radiation intensity
in direction $\Omega$ is
\begin{equation}
	\label{eq:radiation_intensity}
	U(\Omega)
	=
	\left|
	\mathbf{b}^{\mathrm{H}}(\Omega;\mathbf{p})\mathbf{i}
	\right|^2.
\end{equation}
The element factor, polarization factor, and all dimensional normalization
constants are absorbed into $\mathbf{b}$, following the standard far-field
radiation and array-response representation \cite{Balanis2016}. Under the
isolated-element far-field approximation, the response of port $n$ is
\begin{equation}
	\label{eq:simple_array_response}
	b_n(\Omega;\mathbf{p})
	=
	\sqrt{\kappa(\Omega)}
	e^{j\beta\widehat{\mathbf{k}}^{\mathrm{T}}(\Omega)\mathbf{r}_n},
\end{equation}
where $\widehat{\mathbf{k}}(\Omega)$ is the unit propagation vector,
$\beta=2\pi/\lambda$ is the wavenumber, and $\kappa(\Omega)$ contains the
element-pattern normalization \cite{Balanis2016}. For strongly coupled arrays,
the preferred model is an embedded-element-pattern response obtained from
full-wave simulation or measurement: in that case, $\mathbf{b}$ may depend
on all antenna positions, rather than solely through propagation phases
\cite{gupta1983effect,harrington1993field}.

The directivity corresponding to a current vector $\mathbf{i}$ is
\begin{equation}
	\label{eq:directivity}
	D(\Omega;\mathbf{i},\mathbf{p})
	=
	4\pi
	\frac{
		|\mathbf{b}^{\mathrm{H}}(\Omega;\mathbf{p})\mathbf{i}|^2
	}{
		\mathbf{i}^{\mathrm{H}}\mathbf{R}_{\mathrm{rad}}(\mathbf{p})\mathbf{i}
	},
\end{equation}
where the denominator represents the total radiated power under the RMS
phasor convention \cite{Balanis2016}. In the absence of robustness or hardware
constraints, the maximum generalized-Rayleigh-quotient directivity is
\begin{equation}
	\label{eq:unconstrained_directivity}
	D_{\max}(\Omega;\mathbf{p})
	=
	4\pi
	\mathbf{b}^{\mathrm{H}}(\Omega;\mathbf{p})
	\mathbf{R}_{\mathrm{rad}}^{\dagger}(\mathbf{p})
	\mathbf{b}(\Omega;\mathbf{p}),
\end{equation}
provided that $\mathbf{b}(\Omega;\mathbf{p})$ lies in the range of
$\mathbf{R}_{\mathrm{rad}}(\mathbf{p})$. The widely cited $N^2$
superdirectivity scaling is a special asymptotic result for particular
ideal-source array models and should not be interpreted as a universal upper
bound for arbitrary finite dipoles, planar geometries, losses, or scan
directions \cite{Balanis2016,Damico2024}.

For multi-user transmission, let
$\mathbf{h}_k(\mathbf{p})
=
[h_{k,1}(\mathbf{p}),\ldots,h_{k,N}(\mathbf{p})]^\mathrm{T}$
denote the current-to-received-signal channel of user $k$. The received
baseband signal is modeled as
\begin{equation}
	\label{eq:multiuser_signal}
	y_k
	=
	\mathbf{h}_k^\mathrm{H}(\mathbf{p})
	\sum_{j=1}^{K}\mathbf{i}_j s_j
	+
	n_k,
\end{equation}
where $\E[|s_j|^2]=1$, the transmitted symbols are mutually uncorrelated, and
$n_k\sim\CN(0,\sigma_k^2)$. The channel necessarily varies with antenna
positions, since relocating the ports modifies both the propagation phases
and, in general, the embedded radiation responses of the array
\cite{gupta1983effect,Damico2024}. As an example, a single-path far-field
approximation gives
\begin{equation}
	\label{eq:los_channel}
	h_{k,n}(\mathbf{p})
	=
	\alpha_k g_k
	e^{j\beta\widehat{\mathbf{k}}_k^\mathrm{T}\mathbf{r}_n},
\end{equation}
where $\alpha_k$ contains path loss and phase, and $g_k$ captures element-pattern
and polarization factors \cite{Balanis2016}. Near-field, multipath, and
embedded-pattern models can be incorporated without altering the optimization
framework, provided that their position derivatives are available.

\subsection{Superdirectivity Control and Hardware-Robustness Constraints}

Closely spaced arrays may exhibit very small eigenvalues in
$\mathbf{R}_{\mathrm{rad}}$. A beamformer can then exploit weakly radiating
current modes to achieve large directivity, at the expense of large and
cancellation-sensitive currents. This behavior is closely related to the
classical superdirectivity phenomenon in electrically compact arrays
\cite{Balanis2016,Damico2024}. To regulate this effect, we impose
the current-norm constraint
\begin{equation}
	\label{eq:current_constraint}
	\|\mathbf{i}\|_2^2\le\Gamma
\end{equation}
and the source-voltage constraint
\begin{equation}
	\label{eq:voltage_constraint}
	\mathbf{i}^\mathrm{H}\mathbf{Q}_v(\mathbf{p})\mathbf{i}\le V_{\max}^2.
\end{equation}
The minimum-spacing condition in \eqref{eq:position_set} provides an
additional geometric regularization by preventing arbitrarily close antenna
placements.

The current-norm constraint serves as a tractable proxy for hardware
robustness and, through $P_s=R_s\|\mathbf{i}\|_2^2$, admits a direct
source-loss interpretation. It is not, however, mathematically equivalent
to a constraint on antenna quality factor, fractional bandwidth, radiation
efficiency, or matching-network sensitivity. A rigorous $Q$-factor or
bandwidth constraint requires an appropriate stored-energy,
impedance-frequency-derivative, or matching model \cite{Yaghjian2005}.
Accordingly, in this work, the current-norm and source-voltage constraints
are employed as tractable robustness and source-feasibility constraints,
rather than as exact substitutes for bandwidth or $Q$-factor constraints.

\subsection{Position-Differentiable Impedance and Analytical Gradients}

Let
\begin{equation}
	\zeta=p_{n,c},\quad c\in\{x,y\},
\end{equation}
denote one real-valued position coordinate, and define
\begin{equation}
	\mathbf{Z}_{\zeta}
	=
	\frac{\partial\mathbf{Z}}{\partial\zeta}.
\end{equation}
Under the homogeneous distance-dependent impedance model in
\eqref{eq:mutual_impedance}, displacing port $n$ modifies only the $n$th row
and column of $\mathbf{Z}$. Let
\begin{equation}
	d_{mn}=\|\mathbf{p}_m-\mathbf{p}_n\|_2.
\end{equation}
Then, for $m\neq n$,
\begin{equation}
	\label{eq:dZ_position}
	\frac{\partial Z_{mn}}{\partial p_{n,c}}
	=
	\dot{Z}(d_{mn})
	\frac{p_{n,c}-p_{m,c}}{d_{mn}},
\end{equation}
where
\begin{equation}
	\dot{Z}(d)
	=
	\frac{dZ(d)}{dd}.
\end{equation}
Reciprocity yields the corresponding derivative of $Z_{nm}$. The diagonal
derivative vanishes under the translationally invariant free-space model
\cite{harrington1993field,Balanis2016,Damico2024}.

For the mutual-impedance expression in \eqref{eq:mutual_impedance}, the
radial derivative is
\begin{align}
	\label{eq:radial_derivative}
	\dot{Z}(d)
	&=
	\frac{\eta_0}{4\pi}
	\left[
	2\frac{\cos u_0}{u_0}u_0'
	-
	\frac{\cos u_+}{u_+}u_+'
	-
	\frac{\cos u_-}{u_-}u_-'
	\right]
	\nonumber\\
	&\quad
	-j\frac{\eta_0}{4\pi}
	\left[
	2\frac{\sin u_0}{u_0}u_0'
	-
	\frac{\sin u_+}{u_+}u_+'
	-
	\frac{\sin u_-}{u_-}u_-'
	\right],
\end{align}
where
\begin{equation}
	u_0'=\beta,\quad
	u_+'=u_-'
	=
	\frac{\beta d}{\sqrt{d^2+L_{\mathrm d}^{2}}}.
\end{equation}
Equation~\eqref{eq:radial_derivative} follows from the standard identities
\begin{equation}
	\frac{d}{du}\Ci(u)=\frac{\cos u}{u},\quad
	\frac{d}{du}\Si(u)=\frac{\sin u}{u},
\end{equation}
together with the chain rule. The minimum spacing $d_{\min}>0$ prevents
evaluation at the coincident-port singularity $d=0$.

If $\mathbf{R}_{\mathrm{loss}}$ is position independent, then
\begin{subequations}
	\begin{align}
		\label{eq:power_matrix_derivatives}
		\mathbf{R}_{\mathrm{rad},\zeta}
		&=
		\Herm(\mathbf{Z}_{\zeta}),\\
		\mathbf{R}_{\mathrm{acc},\zeta}
		&=
		\Herm(\mathbf{Z}_{\zeta}).
	\end{align}
\end{subequations}
Since $\mathbf{C}=\mathbf{Z}+Z_s\mathbf{I}_N$, we have
$\mathbf{C}_{\zeta}=\mathbf{Z}_{\zeta}$, and hence
\begin{equation}
	\label{eq:Qv_derivative}
	\mathbf{Q}_{v,\zeta}
	=
	\mathbf{Z}_{\zeta}^\mathrm{H}\mathbf{C}
	+
	\mathbf{C}^\mathrm{H}\mathbf{Z}_{\zeta}.
\end{equation}
Consequently, for a fixed current vector,
\begin{subequations}
	\begin{align}
		\label{eq:quadratic_derivatives}
		\frac{\partial}{\partial\zeta}
		\left(
		\mathbf{i}^\mathrm{H}\mathbf{R}_{\mathrm{rad}}\mathbf{i}
		\right)
		&=
		\mathbf{i}^\mathrm{H}
		\mathbf{R}_{\mathrm{rad},\zeta}
		\mathbf{i},
		\\
		\frac{\partial}{\partial\zeta}
		\left(
		\mathbf{i}^\mathrm{H}\mathbf{Q}_v\mathbf{i}
		\right)
		&=
		\mathbf{i}^\mathrm{H}
		\mathbf{Q}_{v,\zeta}
		\mathbf{i}
		\nonumber\\
		&=
		2\operatorname{Re}
		\left\{
		(\mathbf{C}\mathbf{i})^\mathrm{H}
		\mathbf{Z}_{\zeta}\mathbf{i}
		\right\}.
	\end{align}
\end{subequations}

Under the phase-only far-field response model in
\eqref{eq:simple_array_response},
\begin{equation}
	\label{eq:steering_derivative}
	\frac{\partial\mathbf{b}(\Omega;\mathbf{p})}
	{\partial p_{n,c}}
	=
	j\beta\widehat{k}_c(\Omega)
	b_n(\Omega;\mathbf{p})
	\mathbf{e}_n,
\end{equation}
where $\mathbf{e}_n$ is the $n$th canonical vector. This derivative follows
directly from the conventional far-field array-response model \cite{Balanis2016}.
Similarly, under the single-path far-field channel model in
\eqref{eq:los_channel},
\begin{equation}
	\label{eq:channel_derivative}
	\frac{\partial\mathbf{h}_k}
	{\partial p_{n,c}}
	=
	j\beta\widehat{k}_{k,c}h_{k,n}\mathbf{e}_n.
\end{equation}
For a general near-field, multipath, or embedded-pattern model, the complete
derivative $\partial\mathbf{h}_k/\partial p_{n,c}$ must be used. In
particular, it is generally incorrect to differentiate only the impedance
matrix while holding the propagation channel fixed, because antenna
displacement also modifies the propagation response, the embedded radiation
patterns, and the mutual-coupling behavior
\cite{harrington1993field,gupta1983effect,Damico2024}.

\section{Use Case 1: Single-Beam\\ Superdirective Beamforming}
\label{sec:single_beam}

This section applies the current-domain array model to single-beam
superdirective beamforming with movable antenna positions. By normalizing the
desired main-beam response to unity, the problem of maximizing the directivity
in the look direction is equivalently formulated as the minimization of the
total radiated power. To prevent the solution from relying on highly sensitive
superdirective current modes, the formulation includes current-norm,
source-voltage, sidelobe, and geometry constraints. The resulting problem is
nonconvex because the impedance matrix, radiation response, and source-voltage
matrix all depend on the antenna positions. For fixed positions, however, the
current-optimization subproblem is \emph{convex}. This structure motivates a
reduced-gradient alternating method: the optimal current is first computed for a
given geometry, and the antenna positions are then updated using sensitivity
information from the fixed-position subproblem together with a
geometry-feasible line-search step
\cite{Balanis2016,Damico2024,Boyd2004,NocedalWright2006}.

The remainder of this section follows the logic of this decomposition.
Subsection~\ref{subsec:uc1_fixed_current} solves the fixed-position current
subproblem and identifies when a closed-form KKT expression is available.
Subsection~\ref{subsec:uc1_reduced_gradient} derives the reduced position
gradient of the fixed-position optimal value.
Subsection~\ref{subsec:uc1_geometry_step} uses this gradient to construct a
geometry-feasible descent step with minimum-spacing preservation and Armijo
backtracking. Finally,
Subsection~\ref{subsec:uc1_complexity_convergence} analyzes the computational
complexity and local convergence behavior of the single-beam alternating
procedure.

Let $\Omega_0=(\theta_0,\phi_0)$ be the desired beam direction. Let
$\mathcal{Q}=\{1,\ldots,Q_{\mathrm{sl}}\}$ denote the index set of sampled
sidelobe-control directions, and let $\{\Omega_q\}_{q\in\mathcal{Q}}$ be the
corresponding angular grid. We jointly optimize the port-current vector and the
antenna positions:
\begin{subequations}
	\label{eq:uc1_problem}
	\begin{align}
		\underset{\mathbf{i},\,\mathbf{p}}{\operatorname{minimize}}
		\quad&
		\mathbf{i}^H
		\mathbf{R}_{\mathrm{rad}}(\mathbf{p})
		\mathbf{i}
		\label{eq:uc1_obj}
		\\
		\operatorname{subject~to}
		\quad&
		\mathbf{b}_0^H(\mathbf{p})\mathbf{i}=1,
		\label{eq:uc1_mainbeam}
		\\
		&
		\left|
		\mathbf{b}_q^H(\mathbf{p})\mathbf{i}
		\right|
		\le\epsilon_q,~q\in\mathcal{Q},
		\label{eq:uc1_sidelobe}
		\\
		&
		\|\mathbf{i}\|_2^2\le\Gamma,
		\label{eq:uc1_current}
		\\
		&
		\mathbf{i}^H
		\mathbf{Q}_v(\mathbf{p})
		\mathbf{i}
		\le V_{\max}^2,
		\label{eq:uc1_voltage}
		\\
		&
		\mathbf{p}\in\mathcal{P}.
		\label{eq:uc1_geometry}
	\end{align}
\end{subequations}
Here,
$\mathbf{b}_0(\mathbf{p}) = \mathbf{b}(\Omega_0;\mathbf{p})$,
$\mathbf{b}_q(\mathbf{p}) = \mathbf{b}(\Omega_q;\mathbf{p})$,
and $\epsilon_q\ge0$ is the prescribed sidelobe-amplitude bound in direction
$\Omega_q$. The normalization in \eqref{eq:uc1_mainbeam} fixes the numerator
of the directivity in direction $\Omega_0$. Minimizing the radiated power is
therefore equivalent to maximizing the directivity in direction $\Omega_0$,
subject to the current, source-voltage, sidelobe, and geometry constraints
\cite{Balanis2016,Damico2024}.

\begin{algorithm}[t!]
	\caption{Reduced-Gradient Alternating Optimization for Use Case 1:
		Single-Beam Superdirective Beamforming}
	\label{alg:uc1_corrected}
	\begin{algorithmic}[1]
		\STATE Initialize a feasible position vector
		$\mathbf{p}^{(0)}\in\mathcal{P}$
		\STATE Set $t\leftarrow0$
		\REPEAT
		\STATE Build
		$\mathbf{Z}$,
		$\mathbf{R}_{\mathrm{rad}}$,
		$\mathbf{C}$,
		$\mathbf{Q}_v$,
		and the radiation responses
		$\{\mathbf{b}_q\}_{q\in\mathcal{Q}\cup\{0\}}$
		at $\mathbf{p}^{(t)}$
		\STATE Solve the fixed-position convex current problem
		\eqref{eq:uc1_problem} and obtain
		$\mathbf{i}^{(t)}$, $F_1(\mathbf{p}^{(t)})$, and the KKT multipliers
		\STATE Compute the reduced position gradient from
		\eqref{eq:uc1_gradient_expanded}
		\STATE Solve the convex geometry-direction problem
		\eqref{eq:uc1_position_qp}
		\STATE Perform backtracking on $\alpha_t\in(0,1]$; for every trial
		position, re-solve the fixed-position current QCQP
		\STATE Accept the first feasible trial satisfying the Armijo decrease
		condition
		\STATE Update $t\leftarrow t+1$
		\UNTIL{the reduced objective and position vector converge}
		\STATE \textbf{return}
		$\mathbf{i}^{(t)}$ and $\mathbf{p}^{(t)}$
	\end{algorithmic}
\end{algorithm}

\subsection{Fixed-Position Current Subproblem}
\label{subsec:uc1_fixed_current}

For fixed $\mathbf{p}$, problem \eqref{eq:uc1_problem} is a convex
quadratically constrained quadratic program (QCQP). Specifically,
$\mathbf{R}_{\mathrm{rad}}(\mathbf{p})\succeq\mathbf{0}$ and
$\mathbf{Q}_v(\mathbf{p})=\mathbf{C}^H(\mathbf{p})\mathbf{C}(\mathbf{p})
\succeq\mathbf{0}$. The main-beam constraint is a complex affine equality,
equivalent to two real affine equalities. The sidelobe constraints are
second-order cone constraints, and the current and source-voltage constraints
are convex quadratic inequalities. Hence, after the standard real-valued
representation of complex variables, the fixed-position problem can be solved
either as a convex QCQP or as an equivalent second-order cone program (SOCP)
through a conic epigraph reformulation \cite{Boyd2004}.

If the sidelobe constraints are absent, the Karush--Kuhn--Tucker (KKT)
conditions yield the following closed-form current structure:
\begin{equation}
	\label{eq:uc1_closed_form}
	\mathbf{i}^{\star}(\mu,\nu)
	=
	\frac{
		\mathbf{A}^{-1}(\mu,\nu)\mathbf{b}_0
	}{
		\mathbf{b}_0^H
		\mathbf{A}^{-1}(\mu,\nu)
		\mathbf{b}_0
	},
\end{equation}
where
\begin{equation}
	\label{eq:uc1_A}
	\mathbf{A}(\mu,\nu)
	=
	\mathbf{R}_{\mathrm{rad}}
	+
	\mu\mathbf{I}_N
	+
	\nu\mathbf{Q}_v,\quad
	\mu,\nu\ge0.
\end{equation}
The dual variables satisfy the complementary-slackness conditions
\begin{subequations}
	\begin{align}
		\mu
		\left(
		\|\mathbf{i}^{\star}\|_2^2-\Gamma
		\right)&=0,
		\\
		\nu
		\left(
		{\mathbf{i}^{\star}}^H
		\mathbf{Q}_v
		\mathbf{i}^{\star}
		-V_{\max}^2
		\right)&=0.
	\end{align}
\end{subequations}
The pair $(\mu,\nu)$ can be obtained by solving the associated
two-dimensional dual problem, or equivalently by minimizing the negative dual
function over $\mu,\nu\ge0$ \cite{Boyd2004}. A scalar diagonal loading
$\mathbf{R}_{\mathrm{rad}}+\delta\mathbf{I}_N$ is equivalent only to a
current-norm penalty; it does not reproduce the source-voltage constraint
unless $\mathbf{Q}_v$ is proportional to the identity. If $\mathbf{A}$ is
singular, its Moore--Penrose inverse may be used when the normalization remains
feasible, or a strictly positive proximal regularization may be introduced.

When sidelobe constraints are present, the active constraints introduce
additional KKT terms that are not captured by the two-parameter matrix
$\mathbf{A}(\mu,\nu)$ in \eqref{eq:uc1_A}. The closed-form expression in
\eqref{eq:uc1_closed_form} is therefore no longer applicable; instead, the
fixed-position current subproblem is solved directly as a convex QCQP or
through its equivalent SOCP reformulation \cite{Boyd2004}. The proposed
algorithm alternates between an exact current optimization at fixed positions
and a reduced-gradient update of the antenna positions, using a convex inner
approximation of the spacing constraints and a backtracking line search on the
reduced objective, following standard ideas from sequential convex
approximation and line-search optimization \cite{Beck2010,NocedalWright2006}.

Algorithm~\ref{alg:uc1_corrected} should be interpreted as an optimization
method for the reduced objective $F_1(\mathbf{p})$, not as a procedure that
moves the antennas while keeping the previous current vector fixed. At each
position $\mathbf{p}^{(t)}$, the best feasible current vector is first found
by solving the convex current subproblem. The KKT multipliers of that
subproblem are then used to quantify how the optimal value changes when the
antenna positions are perturbed. The geometry-direction problem produces a
feasible local displacement that respects the aperture and a conservative
linearization of the minimum-spacing constraints. Finally, the line search
tests trial positions by re-solving the current subproblem at each trial
geometry. This re-optimization is necessary because
$\mathbf{b}_q(\mathbf{p})$, $\mathbf{R}_{\mathrm{rad}}(\mathbf{p})$, and
$\mathbf{Q}_v(\mathbf{p})$ all change as the antennas move; a current vector
that is feasible at $\mathbf{p}^{(t)}$ need not remain feasible at the new
position.

\subsection{Reduced Position Gradient}
\label{subsec:uc1_reduced_gradient}

Let $
	F_1(\mathbf{p})
	=
	\min_{\mathbf{i}}
	\left\{
	\mathbf{i}^H\mathbf{R}_{\mathrm{rad}}(\mathbf{p})\mathbf{i}:
	\eqref{eq:uc1_mainbeam}
	\text{--}
	\eqref{eq:uc1_voltage}
	\right\}$
denote the optimal value of the fixed-position current subproblem. Let
$\kappa\in\mathbb{C}$ be the multiplier of the main-beam equality,
$\alpha_q\ge0$ the sidelobe multipliers, and $\mu,\nu\ge0$ the current and
source-voltage multipliers. Using the equivalent squared sidelobe constraints
$|\mathbf{b}_q^H\mathbf{i}|^2\le\epsilon_q^2$, the current-subproblem
Lagrangian is
\begin{align}
	\label{eq:uc1_lagrangian}
	&\mathcal{L}_1
	=
	\mathbf{i}^H\mathbf{R}_{\mathrm{rad}}\mathbf{i}
	+
	2\operatorname{Re}
	\left\{
	\kappa^*
	\left(
	\mathbf{b}_0^H\mathbf{i}-1
	\right)
	\right\}
	\nonumber\\
	&+
	\sum_{q\in\mathcal{Q}}
	\alpha_q
	\left(
	|\mathbf{b}_q^H\mathbf{i}|^2-\epsilon_q^2
	\right)
	+
	\mu
	\left(
	\|\mathbf{i}\|_2^2-\Gamma
	\right)
	+
	\nu
	\left(
	\mathbf{i}^H\mathbf{Q}_v\mathbf{i}
	-V_{\max}^2
	\right).
\end{align}
Under local uniqueness and the usual constraint qualification, the envelope
theorem for parametric optimization gives
\begin{equation}
	\label{eq:uc1_reduced_gradient}
	\frac{\partial F_1}{\partial\zeta}
	=
	\left.
	\frac{\partial\mathcal{L}_1}{\partial\zeta}
	\right|_{\mathbf{i}=\mathbf{i}^{\star}},
\end{equation}
where $\zeta$ denotes one real-valued position coordinate
\cite{BonnansShapiro2000}. Expanding the explicit dependence on $\zeta$ yields
\begin{align}
	\label{eq:uc1_gradient_expanded}
	\frac{\partial F_1}{\partial\zeta}
	&=
	{\mathbf{i}^{\star}}^H
	\mathbf{R}_{\mathrm{rad},\zeta}
	\mathbf{i}^{\star}
	\nonumber\\
	&\quad+
	2\operatorname{Re}
	\left\{
	\kappa^*
	\mathbf{b}_{0,\zeta}^H
	\mathbf{i}^{\star}
	\right\}
	\nonumber\\
	&\quad+
	2\sum_{q\in\mathcal{Q}}
	\alpha_q
	\operatorname{Re}
	\left\{
	\left(
	\mathbf{b}_q^H\mathbf{i}^{\star}
	\right)^*
	\mathbf{b}_{q,\zeta}^H\mathbf{i}^{\star}
	\right\}
	\nonumber\\
	&\quad+
	\nu
	{\mathbf{i}^{\star}}^H
	\mathbf{Q}_{v,\zeta}
	\mathbf{i}^{\star}.
\end{align}
The current-norm term does not appear in \eqref{eq:uc1_gradient_expanded}
because it carries no explicit dependence on the antenna positions.

\subsection{Geometry-Feasible Position Update}
\label{subsec:uc1_geometry_step}

Let $\mathbf{d}_n\in\mathbb{R}^2$ be a displacement of antenna $n$, and let
$\mathbf{d}$ stack all displacements. At position iterate $\mathbf{p}^{(t)}$,
a geometry-feasible descent direction is found from
\begin{subequations}
	\label{eq:uc1_position_qp}
	\begin{align}
		\underset{\mathbf{d}}{\operatorname{minimize}}
		\quad&
		\nabla F_1(\mathbf{p}^{(t)})^T\mathbf{d}
		+
		\frac{1}{2\tau_t}\|\mathbf{d}\|_2^2
		\\
		\operatorname{subject~to}
		\quad&
		\mathbf{p}_n^{(t)}+\mathbf{d}_n\in\mathcal{A},
		\quad n=1,\ldots,N,
		\\
		&
		\|\boldsymbol{\delta}_{mn}^{(t)}\|_2^2
		+
		2{\boldsymbol{\delta}_{mn}^{(t)}}^T
		(\mathbf{d}_m-\mathbf{d}_n)
		\ge d_{\min}^2,\quad m<n,
		\\
		&
		\|\mathbf{d}\|_2\le\Delta_t,
	\end{align}
\end{subequations}
where
$\boldsymbol{\delta}_{mn}^{(t)} = \mathbf{p}_m^{(t)} - \mathbf{p}_n^{(t)}.$
The spacing constraint in \eqref{eq:uc1_position_qp} is a convex inner
approximation, obtained from the first-order lower bound of the convex
function $\|\cdot\|_2^2$:
\begin{align}
	\left\|
	\boldsymbol{\delta}_{mn}^{(t)}
	+
	\mathbf{d}_m-\mathbf{d}_n
	\right\|_2^2
	\ge
	\|\boldsymbol{\delta}_{mn}^{(t)}\|_2^2
	+
	2{\boldsymbol{\delta}_{mn}^{(t)}}^T
	(\mathbf{d}_m-\mathbf{d}_n).
\end{align}
Enforcing this linearized lower bound to be at least $d_{\min}^2$ therefore
guarantees that the true pairwise distance constraint is satisfied for the
trial geometry. This conservative convexification is consistent with standard
inner approximation and sequential convex approximation methods
\cite{Boyd2004,Beck2010}.

A backtracking line search is then performed on
\begin{equation}
	\mathbf{p}_{\mathrm{trial}}
	=
	\mathbf{p}^{(t)}
	+
	\alpha_t\mathbf{d}.
\end{equation}
For each trial value of $\alpha_t$, the fixed-position current QCQP is
re-solved at $\mathbf{p}_{\mathrm{trial}}$. The trial step is accepted only if
the current subproblem is feasible and the reduced objective satisfies the
Armijo decrease condition
\begin{equation}
	F_1(\mathbf{p}_{\mathrm{trial}})
	\le
	F_1(\mathbf{p}^{(t)})
	+
	c_{\mathrm A}\alpha_t
	\nabla F_1(\mathbf{p}^{(t)})^T\mathbf{d},
	\quad 0<c_{\mathrm A}<1.
\end{equation}
If this condition fails, $\alpha_t$ is reduced and the trial is repeated. The
accepted step is thus evaluated using the re-optimized current at the new
antenna positions, rather than the current vector from the previous geometry.
This yields a descent step for the reduced problem and avoids accepting a
position update whose associated current subproblem is infeasible
\cite{NocedalWright2006}.

\subsection{Complexity and Convergence Analysis}
\label{subsec:uc1_complexity_convergence}

The computational cost of one outer iteration is dominated by the fixed-position
current solve and the line-search re-solves. Constructing the impedance matrix
$\mathbf{Z}$ requires $\mathcal{O}(N^2)$ operations. Because each position
derivative $\mathbf{Z}_{\zeta}$ has nonzero entries only in one row and one
column, all electromagnetic gradient terms can be accumulated over antenna
pairs, so that computing the full reduced gradient also costs
$\mathcal{O}(N^2)$. By contrast, the fixed-position current subproblem is a
convex QCQP whose dominant cost is $\mathcal{O}(N^3)$ per main solve or dual
evaluation, whether handled through the closed-form KKT structure or a generic
conic solver. The geometry update requires solving a $2N$-variable convex QP
with $\mathcal{O}(N^2)$ linearized spacing constraints, after which several
backtracking trials may each require re-solving the current subproblem. The
practical runtime is therefore governed primarily by repeated current-subproblem
solves inside the outer alternating loop, rather than by the gradient or
geometry-update computations alone.

Regarding convergence, the current subproblem is solved globally at every fixed
geometry because it is convex. The position update performs a reduced-gradient
step on the optimal-value function $F_1(\mathbf{p})$, using envelope-theorem
sensitivities together with a first-order inner approximation of the spacing
constraints. Each accepted position step is validated by re-solving the current
QCQP and enforcing an Armijo decrease condition on the re-optimized reduced
objective; consequently, the outer iterations produce a monotone nonincreasing
sequence of objective values. Under standard assumptions of local smoothness
of the position-dependent matrices and steering responses, compactness of the
feasible geometry set, and appropriate constraint qualifications for the
parametric current subproblem, the alternating procedure converges to a
stationary point of the reduced single-beam problem. Because the overall
problem is nonconvex in the antenna positions, this stationary point is
generally local rather than globally optimal.

\section{Use Case 2: Multi-User Weighted Sum-Rate}
\label{sec:multiuser}

This section considers downlink multi-user transmission from the planar FAA to
$K$ single-antenna users. In contrast to the single-beam design in
Section~\ref{sec:single_beam}, where one normalized beam response is enforced,
the multi-user problem optimizes one current vector per user data stream.
Specifically, $\mathbf{i}_k$ is the current-domain beamforming vector for user
$k$, while the collection
$\mathbf{I}_{\mathrm{p}}=[\mathbf{i}_1,\ldots,\mathbf{i}_K]$
is the corresponding current-domain linear precoder. In the multi-user setting,
``beamforming'' refers to the spatial field generated by a single stream, whereas
``precoding'' refers to the joint design of all stream-dependent current vectors
that balances desired-signal enhancement against inter-user interference
suppression \cite{Shen2018,Shi2011}. The objective is to maximize the weighted
sum spectral efficiency under accepted-power, current-norm, source-voltage, and
geometry constraints. Because the user channels, impedance matrix,
accepted-power matrix, and source-voltage matrix all depend on the antenna
positions, the resulting problem couples communication-theoretic precoding with
position-dependent electromagnetic constraints \cite{Damico2024}.

The remainder of this section follows the same fixed-position/geometry
decomposition as Section~\ref{sec:single_beam}.
Subsection~\ref{subsec:uc2_fixed_fp} develops the fixed-position
fractional-programming (FP) precoder update.
Subsection~\ref{subsec:uc2_position_gradient} derives the reduced position
gradient of the locally re-optimized weighted sum rate.
Subsection~\ref{subsec:uc2_geometry_step} constructs a geometry-feasible ascent
step and accepts trial positions only after re-solving the fixed-position FP
problem. Finally, Subsection~\ref{subsec:uc2_complexity_convergence} analyzes the
computational complexity and local convergence properties of the FP--geometry
alternating algorithm.

Let $\mathcal{K}=\{1,\ldots,K\}$ denote the user index set. For user
$k\in\mathcal{K}$, let $\mathbf{h}_k(\mathbf{p})\in\mathbb{C}^{N}$ be the
current-to-received-signal channel, let $\sigma_k^2$ be the noise variance, and
let $\omega_k>0$ be the priority weight. The transmitted current vector is
$\sum_{j=1}^{K}\mathbf{i}_j s_j$, where the data symbols satisfy
$\mathbb{E}[|s_j|^2]=1$ and are mutually uncorrelated. The
signal-to-interference-plus-noise ratio (SINR) of user $k$ is
\begin{equation}
	\label{eq:sinr}
	\gamma_k
	=
	\frac{
		|\mathbf{h}_k^H(\mathbf{p})\mathbf{i}_k|^2
	}{
		\displaystyle
		\sum_{j\neq k}
		|\mathbf{h}_k^H(\mathbf{p})\mathbf{i}_j|^2
		+
		\sigma_k^2
	}.
\end{equation}
The joint current and position problem is
\begin{subequations}
	\label{eq:uc2_problem}
	\begin{align}
		\underset{
			\{\mathbf{i}_k\}_{k=1}^{K},\,\mathbf{p}
		}{\operatorname{maximize}}
		\quad&
		R_{\mathrm{sum}}
		=
		\sum_{k=1}^{K}
		\omega_k\log_2(1+\gamma_k)
		\label{eq:uc2_obj}
		\\
		\operatorname{subject~to}
		\quad&
		\sum_{k=1}^{K}
		\mathbf{i}_k^H
		\mathbf{R}_{\mathrm{acc}}(\mathbf{p})
		\mathbf{i}_k
		\le P_{\max},
		\label{eq:uc2_power}
		\\
		&
		\sum_{k=1}^{K}
		\|\mathbf{i}_k\|_2^2
		\le\Gamma,
		\label{eq:uc2_current}
		\\
		&
		\sum_{k=1}^{K}
		\mathbf{i}_k^H
		\mathbf{Q}_v(\mathbf{p})
		\mathbf{i}_k
		\le V_{\max}^2,
		\label{eq:uc2_voltage}
		\\
		&
		\mathbf{p}\in\mathcal{P}.
		\label{eq:uc2_geometry}
	\end{align}
\end{subequations}
Here, $P_{\max}$ is the accepted-power budget, $\Gamma$ is the total
current-norm budget, and $V_{\max}$ is the source-voltage budget. The feasible
position set $\mathcal{P}$ is defined in \eqref{eq:position_set}. The
accepted-power and source-voltage constraints are written as sums over $k$
because the data streams are mutually uncorrelated. The accepted-power
constraint uses $\mathbf{R}_{\mathrm{acc}}$ rather than
$\mathbf{R}_{\mathrm{rad}}$ so that antenna dissipation is included. A purely
radiated-power budget can instead be imposed by replacing
$\mathbf{R}_{\mathrm{acc}}$ with $\mathbf{R}_{\mathrm{rad}}$.

\subsection{Fixed-Position Fractional-Programming Update}
\label{subsec:uc2_fixed_fp}

For fixed $\mathbf{p}$, the weighted sum-rate problem remains nonconvex because
each user's desired-signal and interference terms are coupled through the
precoding currents. We address this fixed-position subproblem using the
Lagrangian-dual and quadratic transforms of FP \cite{Shen2018}. This FP
procedure should be interpreted as a monotonic block-coordinate method that
converges to a stationary point of the fixed-position weighted sum-rate problem,
rather than as a global optimization method for the nonconvex sum-rate objective
\cite{Shen2018,Shi2011}. Natural logarithms are used in the FP derivation;
multiplication by $1/\ln2$ recovers rates in bits/s/Hz and does not change the
current maximizer. When the KKT multipliers are reused in the reduced-gradient
formula, their scaling must be consistent with the logarithm base used in the
Lagrangian.

For notational compactness, the explicit dependence on $\mathbf{p}$ is omitted
within this subsection. Define the total received power at user $k$ as
\begin{equation}
	\label{eq:total_received_power}
	T_k
	=
	\sum_{j=1}^{K}
	|\mathbf{h}_k^H\mathbf{i}_j|^2
	+
	\sigma_k^2.
\end{equation}
Given the current vectors, the FP auxiliary variables are updated as
\begin{subequations}
	\begin{align}
		\label{eq:beta_update}
		\beta_k&=\gamma_k,\\
		\label{eq:xi_update}
		\xi_k
		&=
		\frac{
			\sqrt{\omega_k(1+\beta_k)}
			\mathbf{h}_k^H\mathbf{i}_k
		}{
			T_k
		},\quad k\in\mathcal{K}.
	\end{align}
\end{subequations}
Here, $\beta_k$ and $\xi_k$ are scalar FP auxiliary variables; $\beta_k$ should
not be confused with the wavenumber $\beta=2\pi/\lambda$ used in the array
response model. Up to terms independent of the current vectors, the transformed
objective is
\begin{align}
	\label{eq:fp_objective}
	f_{\mathrm{FP}}
	&=
	\sum_{k=1}^{K}
	\omega_k
	\left[
	\ln(1+\beta_k)-\beta_k
	\right]
	\nonumber\\
	&\quad
	+
	\sum_{k=1}^{K}
	\left[
	2
	\sqrt{\omega_k(1+\beta_k)}
	\operatorname{Re}
	\left\{
	\xi_k^*
	\mathbf{h}_k^H\mathbf{i}_k
	\right\}
	-
	|\xi_k|^2T_k
	\right].
\end{align}
For fixed $\{\beta_k,\xi_k\}_{k=1}^{K}$, define
\begin{subequations}
	\begin{align}
		\label{eq:G_matrix}
		\mathbf{G}
		&=
		\sum_{\ell=1}^{K}
		|\xi_\ell|^2
		\mathbf{h}_\ell\mathbf{h}_\ell^H,\\
		\label{eq:c_vector}
		\mathbf{c}_k
		&=
		\sqrt{\omega_k(1+\beta_k)}
		\xi_k\mathbf{h}_k.
	\end{align}
\end{subequations}
Let $\lambda,\mu,\nu\ge0$ be the multipliers of \eqref{eq:uc2_power},
\eqref{eq:uc2_current}, and \eqref{eq:uc2_voltage}, respectively. For fixed
auxiliary variables, the current update solves a convex quadratic maximization
problem with convex quadratic constraints. Its KKT update has the form
\begin{subequations}
	\begin{align}
		\label{eq:fp_current_update}
		\mathbf{i}_k
		&=
		\mathbf{M}^{-1}\mathbf{c}_k,\quad k\in\mathcal{K},
		\\
		\label{eq:M_matrix}
		\mathbf{M}
		&=
		\mathbf{G}
		+
		\lambda\mathbf{R}_{\mathrm{acc}}
		+
		\mu\mathbf{I}_N
		+
		\nu\mathbf{Q}_v,
	\end{align}
\end{subequations}
whenever $\mathbf{M}\succ\mathbf{0}$. If $\mathbf{M}$ is singular, the
minimum-norm KKT solution can be computed with the Moore--Penrose inverse,
provided that $\mathbf{c}_k$ lies in the range of $\mathbf{M}$.

For a guaranteed unique numerical update, a proximal minorization term
\begin{equation}
	-\tau_i
	\sum_{k=1}^{K}
	\|
	\mathbf{i}_k-\mathbf{i}_k^{(q)}
	\|_2^2,\quad\tau_i>0,
\end{equation}
may be added at inner FP iteration $q$. In this case,
\eqref{eq:fp_current_update} is replaced by
\begin{equation}
	\label{eq:proximal_fp_update}
	\mathbf{i}_k
	=
	\left(
	\mathbf{M}+\tau_i\mathbf{I}_N
	\right)^{-1}
	\left(
	\mathbf{c}_k+\tau_i\mathbf{i}_k^{(q)}
	\right).
\end{equation}
Unlike an unexplained numerical diagonal guard, this proximal term is an
explicit part of the block-minorization algorithm and is tight at the current
iterate.

For the non-proximal update, the dual variables are obtained jointly by
minimizing the convex dual function
\begin{align}
	\label{eq:dual_function}
	g(\lambda,\mu,\nu)
	&=
	\lambda P_{\max}
	+
	\mu\Gamma
	+
	\nu V_{\max}^2
	\nonumber\\
	&\quad
	+
	\sum_{k=1}^{K}
	\mathbf{c}_k^H
	\mathbf{M}^{-1}
	\mathbf{c}_k
	+
	\mathrm{constant},
\end{align}
over $\lambda,\mu,\nu\ge0$ \cite{Boyd2004}. The proximal version is obtained by
replacing $\mathbf{M}$ and $\mathbf{c}_k$ according to
\eqref{eq:proximal_fp_update}, together with the corresponding constant terms.
For the non-proximal case, the derivatives of \eqref{eq:dual_function} are
\begin{subequations}
	\begin{align}
		\label{eq:dual_gradients}
		\frac{\partial g}{\partial\lambda}
		&=
		P_{\max}
		-
		\sum_{k=1}^{K}
		\mathbf{i}_k^H
		\mathbf{R}_{\mathrm{acc}}
		\mathbf{i}_k,
		\\
		\frac{\partial g}{\partial\mu}
		&=
		\Gamma
		-
		\sum_{k=1}^{K}
		\|\mathbf{i}_k\|_2^2,
		\\
		\frac{\partial g}{\partial\nu}
		&=
		V_{\max}^2
		-
		\sum_{k=1}^{K}
		\mathbf{i}_k^H
		\mathbf{Q}_v
		\mathbf{i}_k.
	\end{align}
\end{subequations}
A projected Newton, projected quasi-Newton, or other convex dual solver can be
used. Because all three resource consumptions depend simultaneously on all three
dual variables through $\mathbf{M}^{-1}$, a nested scalar bisection is not
generally an exact solution method unless additional monotonic-separability
properties are established.

\subsection{Reduced Position Gradient}
\label{subsec:uc2_position_gradient}

For fixed current vectors, define
\begin{subequations}
	\begin{align}
		z_{kj}
		&=
		\mathbf{h}_k^H\mathbf{i}_j,\\
		D_k
		&=
		\sum_{j\neq k}|z_{kj}|^2+\sigma_k^2,\\
		T_k
		&=
		D_k+|z_{kk}|^2.
	\end{align}
\end{subequations}
For a real position coordinate $\zeta$, let
$z_{kj,\zeta} = \mathbf{h}_{k,\zeta}^H\mathbf{i}_j$ with
$\mathbf{h}_{k,\zeta} = \partial\mathbf{h}_k/\partial\zeta$.
Then
\begin{align}
	\label{eq:rate_position_derivative}
	\frac{\partial R_{\mathrm{sum}}}{\partial\zeta}
	&=
	\frac{2}{\ln2}
	\sum_{k=1}^{K}
	\omega_k
	\Bigg[
	\frac{
		\displaystyle
		\sum_{j=1}^{K}
		\operatorname{Re}
		\left\{
		z_{kj}^*z_{kj,\zeta}
		\right\}
	}{
		T_k
	}
	\nonumber\\
	&\hspace{36mm}
	-
	\frac{
		\displaystyle
		\sum_{j\neq k}
		\operatorname{Re}
		\left\{
		z_{kj}^*z_{kj,\zeta}
		\right\}
	}{
		D_k
	}
	\Bigg].
\end{align}

Let $\lambda,\mu,\nu$ denote the KKT multipliers of the fixed-position current
problem, with the same logarithm-base convention as the rate objective in
\eqref{eq:uc2_obj}. Its constrained-maximization Lagrangian is
\begin{align}
	\label{eq:uc2_lagrangian}
	\mathcal{L}_2
	&=
	R_{\mathrm{sum}}
	-
	\lambda
	\left(
	\sum_{k=1}^{K}
	\mathbf{i}_k^H
	\mathbf{R}_{\mathrm{acc}}
	\mathbf{i}_k
	-
	P_{\max}
	\right)
	\nonumber\\
	&\quad
	-
	\mu
	\left(
	\sum_{k=1}^{K}
	\|\mathbf{i}_k\|_2^2
	-
	\Gamma
	\right)
	\nonumber\\
	&\quad
	-
	\nu
	\left(
	\sum_{k=1}^{K}
	\mathbf{i}_k^H
	\mathbf{Q}_v
	\mathbf{i}_k
	-
	V_{\max}^2
	\right).
\end{align}
Under local regularity of the fixed-position stationary point, the
envelope-theorem sensitivity of the local reduced objective follows from the
explicit position dependence of the Lagrangian \cite{BonnansShapiro2000}. The
reduced ascent gradient with respect to $\zeta$ is therefore
\begin{align}
	\label{eq:uc2_reduced_gradient}
	\frac{\partial\mathcal{L}_2}{\partial\zeta}
	&=
	\frac{\partial R_{\mathrm{sum}}}{\partial\zeta}
	-
	\lambda
	\sum_{k=1}^{K}
	\mathbf{i}_k^H
	\mathbf{R}_{\mathrm{acc},\zeta}
	\mathbf{i}_k
	\nonumber\\
	&\quad
	-
	\nu
	\sum_{k=1}^{K}
	\mathbf{i}_k^H
	\mathbf{Q}_{v,\zeta}
	\mathbf{i}_k.
\end{align}
The current-norm term does not appear in \eqref{eq:uc2_reduced_gradient} because
it has no explicit position dependence. Both the accepted-power derivative and
the source-voltage derivative must be retained in the position update. In
addition, the channel derivative $\mathbf{h}_{k,\zeta}$ in
\eqref{eq:rate_position_derivative} must be evaluated using the chosen
propagation model; for near-field, multipath, or embedded-pattern models, it is
generally insufficient to differentiate only the phase term or only the
impedance matrix.

\subsection{Geometry-Feasible Position Update}
\label{subsec:uc2_geometry_step}

At outer iteration $t$, let $\mathbf{d}_n\in\mathbb{R}^2$ denote the
displacement of antenna $n$, and let $\mathbf{d}$ stack all antenna
displacements. A geometry-feasible ascent direction is obtained from
\begin{subequations}
	\label{eq:uc2_position_qp}
	\begin{align}
		\underset{\mathbf{d}}{\operatorname{maximize}}
		\quad&
		\nabla_{\mathbf{p}}\mathcal{L}_2^\mathrm{T}\mathbf{d}
		-
		\frac{1}{2\tau_t}\|\mathbf{d}\|_2^2
		\\
		\operatorname{subject~to}
		\quad&
		\mathbf{p}_n^{(t)}+\mathbf{d}_n\in\mathcal{A},
		\quad n=1,\ldots,N,
		\\
		&
		\|\boldsymbol{\delta}_{mn}^{(t)}\|_2^2
		+
		2{\boldsymbol{\delta}_{mn}^{(t)}}^T
		(\mathbf{d}_m-\mathbf{d}_n)
		\ge d_{\min}^2,
		\quad m<n,
		\\
		&
		\|\mathbf{d}\|_2\le\Delta_t,
	\end{align}
\end{subequations}
where $\tau_t>0$ is the proximal stepsize, $\Delta_t>0$ is the trust-region
radius, and
$\boldsymbol{\delta}_{mn}^{(t)} = \mathbf{p}_m^{(t)} - \mathbf{p}_n^{(t)}$.
The spacing constraint in \eqref{eq:uc2_position_qp} is an inner convex
approximation of the minimum-spacing constraint, obtained from the first-order
lower bound of the convex function $\|\cdot\|_2^2$:
\begin{align}
	\left\|
	\boldsymbol{\delta}_{mn}^{(t)}
	+
	\mathbf{d}_m-\mathbf{d}_n
	\right\|_2^2
	\ge
	\|\boldsymbol{\delta}_{mn}^{(t)}\|_2^2
	+
	2{\boldsymbol{\delta}_{mn}^{(t)}}^T
	(\mathbf{d}_m-\mathbf{d}_n).
\end{align}
Enforcing this linearized lower bound to be at least $d_{\min}^2$ therefore
guarantees that the true pairwise spacing constraint is satisfied for the trial
geometry. This conservative convexification is consistent with standard inner
approximation and sequential convex approximation methods \cite{Boyd2004,Beck2010}.

A backtracking line search is then performed on
\begin{equation}
	\mathbf{p}_{\mathrm{trial}}
	=
	\mathbf{p}^{(t)}
	+
	\alpha_t\mathbf{d},
	\quad \alpha_t\in(0,1].
\end{equation}
For each trial value of $\alpha_t$, the fixed-position FP stage is rerun at
$\mathbf{p}_{\mathrm{trial}}$, initialized from the previous current vectors
after feasible scaling if necessary. Let $R_{\mathrm{sum}}^{\mathrm{trial}}$
denote the feasible weighted sum rate returned by the FP stage at the trial
geometry. The trial step is accepted only if
\begin{equation}
	R_{\mathrm{sum}}^{\mathrm{trial}}
	\ge
	R_{\mathrm{sum}}^{(t)}
	+
	c_{\mathrm A}\alpha_t
	\nabla_{\mathbf{p}}\mathcal{L}_2^\mathrm{T}\mathbf{d},
	\quad 0<c_{\mathrm A}<1,
\end{equation}
where $c_{\mathrm A}$ is the Armijo parameter. If this condition fails,
$\alpha_t$ is reduced and the trial is repeated \cite{NocedalWright2006}. This
line-search test uses the re-optimized feasible precoder at the new antenna
positions, thereby avoiding acceptance of a geometry update solely because the
old current vectors appear favorable under a local linear model, even though
those old currents may no longer be feasible or well adapted after the channel,
impedance, accepted-power, and source-voltage matrices have changed.

The complete procedure is summarized in Algorithm~\ref{alg:uc2_corrected}. The
algorithm has two nested levels. The inner level fixes the antenna positions and
updates the current-domain precoder using FP until the fixed-position weighted
sum rate ceases to improve. The outer level treats the resulting stationary
sum-rate value as a local reduced objective in the antenna positions, computes a
position gradient from the channel and electromagnetic derivatives, and updates
the geometry through a conservative spacing-aware line search.

\begin{algorithm}[t]
	\caption{FP and Reduced-Gradient Position Optimization for Use Case 2: Multi-User Weighted Sum-Rate}
	\label{alg:uc2_corrected}
	\begin{algorithmic}[1]
		\STATE Initialize
		$\mathbf{p}^{(0)}\in\mathcal{P}$
		and nonzero feasible current vectors
		$\{\mathbf{i}_k^{(0)}\}_{k=1}^{K}$
		\STATE Set $t\leftarrow0$
		\REPEAT
		\STATE Build
		$\mathbf{Z}$,
		$\mathbf{R}_{\mathrm{acc}}$,
		$\mathbf{C}$,
		$\mathbf{Q}_v$,
		and $\{\mathbf{h}_k\}_{k=1}^{K}$
		at $\mathbf{p}^{(t)}$
		\STATE \textbf{Fixed-position FP stage:}
		\REPEAT
		\STATE Update $\beta_k$ from \eqref{eq:beta_update},
		$k\in\mathcal{K}$
		\STATE Update $\xi_k$ from \eqref{eq:xi_update},
		$k\in\mathcal{K}$
		\STATE Jointly solve the convex dual problem
		\eqref{eq:dual_function}
		for $\lambda,\mu,\nu$
		\STATE Update $\{\mathbf{i}_k\}_{k=1}^{K}$ using
		\eqref{eq:fp_current_update} or
		\eqref{eq:proximal_fp_update}
		\UNTIL{the fixed-position FP objective converges}
		\STATE Store the feasible weighted sum rate
		$R_{\mathrm{sum}}^{(t)}$ and the final KKT multipliers
		\STATE Compute the reduced position gradient from
		\eqref{eq:rate_position_derivative} and
		\eqref{eq:uc2_reduced_gradient}
		\STATE Solve the geometry-direction problem
		\eqref{eq:uc2_position_qp}
		\STATE Perform backtracking on the position step; for every trial
		position, rerun the fixed-position FP stage initialized from the
		previous currents after feasible scaling if necessary
		\STATE Accept the first feasible trial satisfying the actual weighted
		sum-rate Armijo condition
		\STATE Update $t\leftarrow t+1$
		\UNTIL{the weighted sum rate and position vector converge}
		\STATE \textbf{return}
		$\{\mathbf{i}_k^{(t)}\}_{k=1}^{K}$ and $\mathbf{p}^{(t)}$
	\end{algorithmic}
\end{algorithm}

\subsection{Complexity and Convergence Analysis}
\label{subsec:uc2_complexity_convergence}

Electromagnetic preprocessing costs $\mathcal{O}(N^2)$, while gradient
evaluation scales as $\mathcal{O}(K^2N)$ owing to pairwise user interactions
across all desired-signal and interference terms. The dominant computational
cost comes from the fixed-position FP stage. Each inner FP iteration forms the
$N\times N$ matrix $\mathbf{M}$ in \eqref{eq:M_matrix} at cost
$\mathcal{O}(KN^2)$, followed by an $N\times N$ factorization costing
$\mathcal{O}(N^3)$ and $K$ associated linear solves. One FP step therefore
costs $\mathcal{O}(N^3+KN^2)$, multiplied by the number of FP iterations and
the cost of the inner convex dual solver for $(\lambda,\mu,\nu)$. The geometry
update solves a convex QP with $2N$ variables and $\mathcal{O}(N^2)$ linearized
spacing constraints, after which several backtracking trials may each require
rerunning the fixed-position FP stage. The practical runtime is consequently
governed primarily by repeated fixed-position FP solves inside the outer
geometry loop rather than by the gradient or geometry-update computations alone.

For every fixed geometry, the FP iterations are monotonic and converge to a
stationary point of the fixed-position weighted sum-rate problem under exact
inner solves \cite{Shen2018,Shi2011}. The outer position update applies a
reduced-gradient ascent step based on envelope-theorem sensitivities, together
with a first-order inner approximation of the spacing constraints. Each accepted
position step is validated by rerunning the fixed-position FP stage at the trial
geometry and enforcing an Armijo increase condition on the re-optimized feasible
weighted sum rate, so the outer iterations produce a monotone nondecreasing
sequence of locally optimized sum-rate values. Under standard assumptions---smooth
position-dependent channels and electromagnetic matrices, compactness of the
feasible geometry set, sufficient accuracy of the inner FP solves, and
appropriate local regularity of the fixed-position stationary solutions---the
alternating FP--geometry procedure converges to a stationary point of the overall
nonconvex problem. Because the problem remains nonconvex in both the multi-user
interference coupling and the antenna positions, global optimality is not
guaranteed.

\begin{figure*}[t!]
	\centering
	\begin{subfigure}[t]{\columnwidth}
		\centering
		\includegraphics[width=0.9\linewidth]{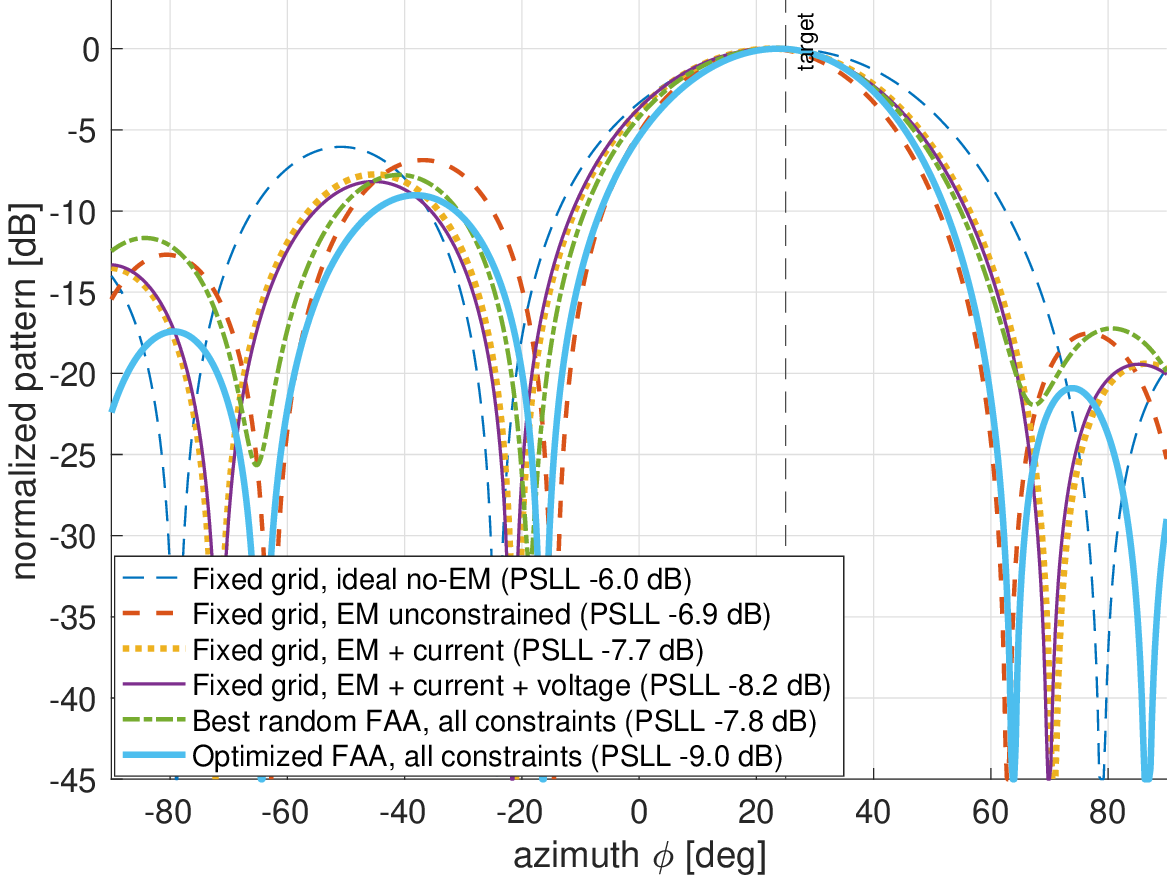}
		\caption{Normalized azimuth beam patterns and PSLL comparison.}
		\label{fig:beam_pattern_2d}
	\end{subfigure}
	\hfill
	\begin{subfigure}[t]{\columnwidth}
		\centering
		\includegraphics[width=0.9\linewidth]{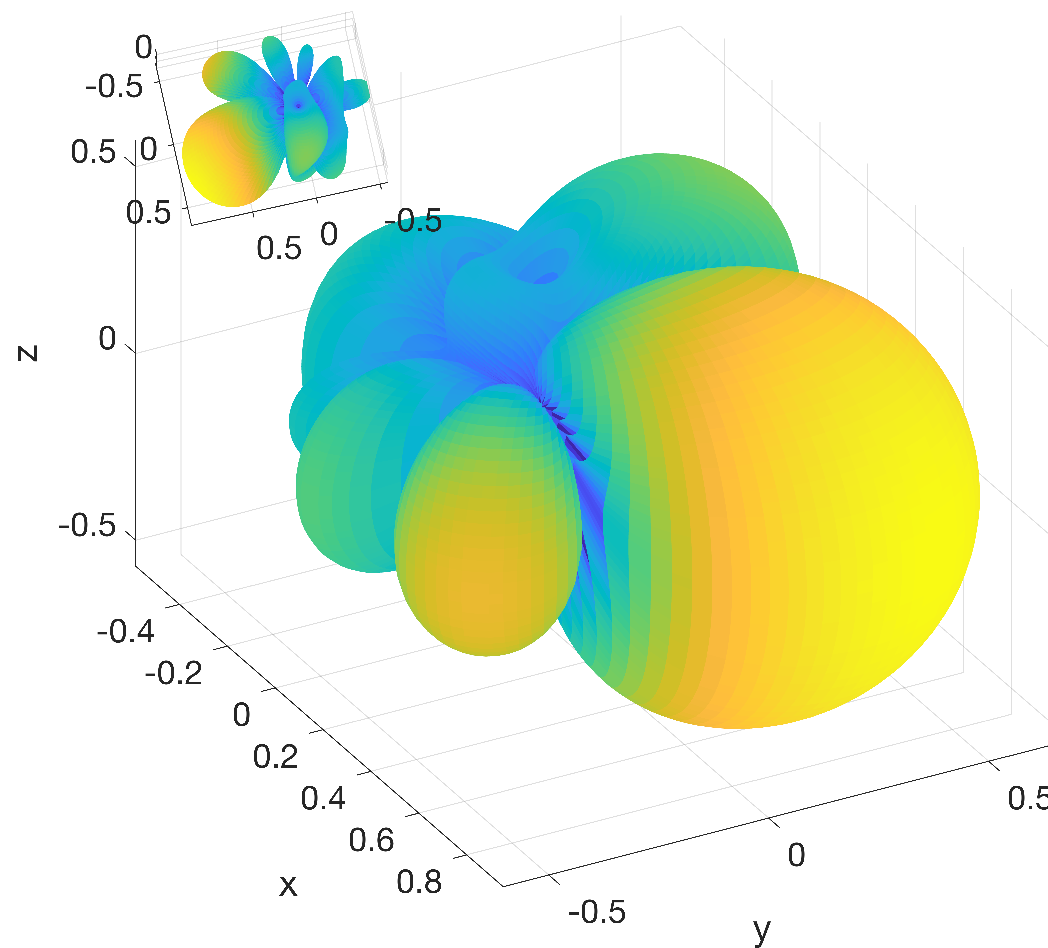}
		\caption{3-D normalized beam pattern of the optimized FAA.}
		\label{fig:beam_pattern_3d}
	\end{subfigure}
	\caption{Single-beam radiation performance. 
		(a) Normalized azimuth beam patterns of different benchmark schemes, where the corresponding PSLL values are shown in the legend. 
		(b) 3-D normalized beam pattern of the optimized FAS under the EM-aware current and voltage constraints.}
	\label{fig:beam_pattern}
	\vspace{-6mm}
\end{figure*}
\begin{figure*}[t!]
	\centering
	\begin{subfigure}[t]{0.49\linewidth}
		\centering
		\includegraphics[width=\linewidth]{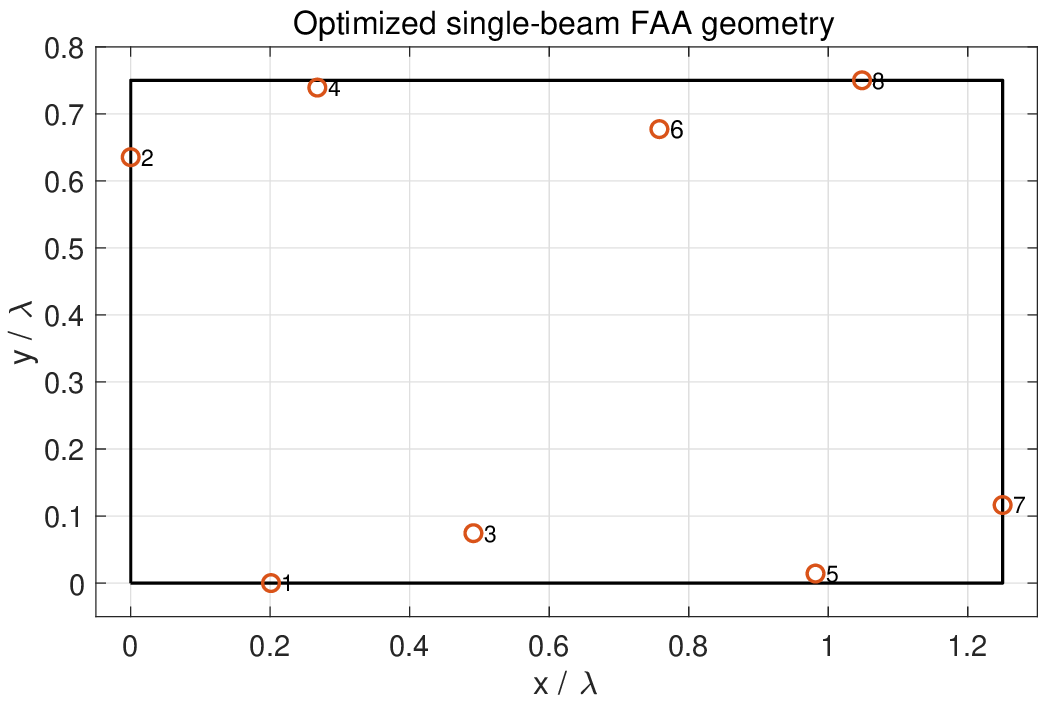}
		\caption{Single-beam case.}
		\label{fig:single_beam_geometry}
	\end{subfigure}
	\hfill
	\begin{subfigure}[t]{0.49\linewidth}
		\centering
		\includegraphics[width=\linewidth]{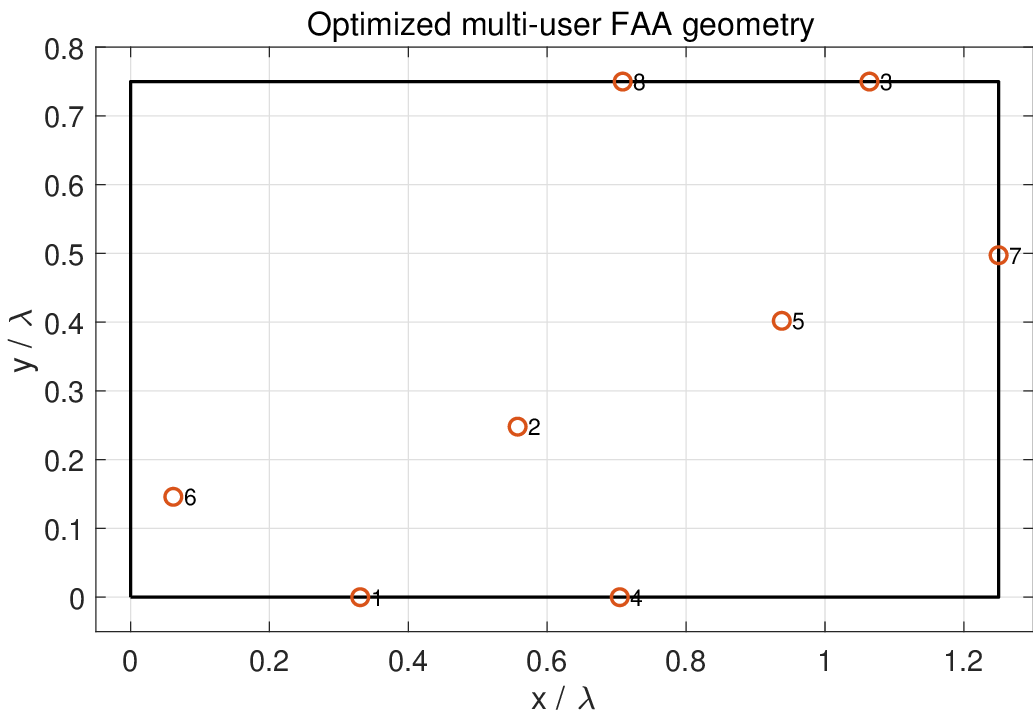}
		\caption{Multi-user case.}
		\label{fig:multi_user_geometry}
	\end{subfigure}
	\caption{Optimized FAA port locations under different objectives. 
		(a) Optimized geometry for the single-beam case, where the objective is to form a beam toward the prescribed target direction while suppressing sidelobes. 
		(b) Optimized geometry for the multi-user case, where the objective is EM-aware weighted sum-rate maximization with multi-user linear precoding.}
	\label{fig:optimized_geometry}
		\vspace{-6mm}
\end{figure*}

\section{Simulation Results}
All simulations are conducted using the EM-aware FAS model with a normalized
wavelength $\lambda=1$. The considered planar FAA consists of $N=8$ configurable ports
deployed within a rectangular aperture of size
$1.25\lambda \times 0.75\lambda$, subject to a minimum inter-port spacing
constraint $d_{\min}=0.2\lambda$. Each port is modeled as a half-wavelength
dipole with length $L_{\mathrm d}=\lambda/2$, self-impedance
$Z_{\mathrm{self}}=73.1+j42.5~\Omega$, port-loss resistance
$R_{\ell}=1~\Omega$, source impedance
$Z_s=R_s+jX_s=50+j0~\Omega$, and reference impedance
$Z_{\mathrm{ref}}=50~\Omega$. The loss-resistance matrix is
$\mathbf{R}_{\mathrm{loss}}=R_{\ell}\mathbf{I}_N$, and the individual dipole
radiation pattern is incorporated into the array response. The port coordinates
are restricted to $0\le x_n\le 1.25\lambda$, $0\le y_n\le 0.75\lambda$, and
$\|\mathbf{p}_n-\mathbf{p}_m\|_2\ge 0.2\lambda$ for $n\neq m$.

For the single-beam experiment, the target direction is set to
$(\theta_0,\phi_0)=(90^\circ,25^\circ)$. The azimuth pattern is sampled over
$[-90^\circ,90^\circ]$ using 721 grid points, and the peak sidelobe level (PSLL)
is evaluated by excluding a $\pm 10^\circ$ main-lobe region around the target
direction. The single-beam optimization enforces the unit-response constraint
$\mathbf{b}^{H}(\Omega_0;\mathbf{p})\mathbf{i}=1$. In the constrained
single-beam cases, the current and voltage budgets are $\Gamma=0.15$ and
$V_{\max}^2=3097$, respectively,
i.e., $\|\mathbf{i}\|_2^2\le 0.150$ and
$\mathbf{i}^{H}\mathbf{Q}_v(\mathbf{p})\mathbf{i}\le 3097$.
Six schemes are compared, corresponding to the legend entries in
Fig.~\ref{fig:beam_pattern}(a):

\begin{enumerate}
	\item {\bf(Fixed grid, ideal no-EM)} is a conventional fixed-array benchmark
	in which the excitation design ignores EM coupling.
	\item {\bf(Fixed grid, EM unconstrained)} adopts the same fixed geometry but
	incorporates the EM-aware radiation model without current or voltage constraints.
	\item {\bf(Fixed grid, EM + current)} further imposes
	$\|\mathbf{i}\|_2^2\le 0.15$.
	\item {\bf(Fixed grid, EM + current + voltage)} jointly enforces
	$\|\mathbf{i}\|_2^2\le 0.15$ and
	$\mathbf{i}^{H}\mathbf{Q}_v(\mathbf{p})\mathbf{i}\le 3097$.
	\item {\bf(Best random FAS, all constraints)} selects the best feasible
	geometry from random FAS realizations under all EM-aware constraints.
	\item {\bf(Optimized FAS, all constraints)} is the proposed scheme, jointly
	optimizing port positions and excitation currents under all physical
	constraints.
\end{enumerate}

For the multi-user experiment, we consider $K=4$ users. Each user channel
comprises four paths with nominal azimuth directions
$[-45^\circ,-10^\circ,25^\circ,60^\circ]$, angular spread $10^\circ$, and path
gains $[0,-3,-6,-9]$ dB. The noise variance is $10^{-3}$, and equal user weights
are adopted. The all-constraint multi-user design uses the accepted-power,
current, and voltage budgets $P_{\max}=1$, $\Gamma=0.03621$, and
$V_{\max}^2=443.4$, namely
$\sum_{k=1}^{K}\mathbf{i}_{k}^{H}\mathbf{R}_{\mathrm{acc}}(\mathbf{p})\mathbf{i}_{k}\le 1$,
$\sum_{k=1}^{K}\|\mathbf{i}_{k}\|_2^2\le 0.03621$, and
$\sum_{k=1}^{K}\mathbf{i}_{k}^{H}\mathbf{Q}_{v}(\mathbf{p})\mathbf{i}_{k}\le 443.4$.
The multi-user benchmarks include the fixed-grid design, the best random FAS, and
the optimized FAS. The fixed-grid FP baselines successively enable the
accepted-power-only, accepted-power-plus-current, and full constraint sets. All
three fair-comparison schemes share the same EM-aware constraints, so that their
comparison isolates the benefit of port-position optimization.

\begin{figure}[t!]
	\centering
	\includegraphics[width=\columnwidth]{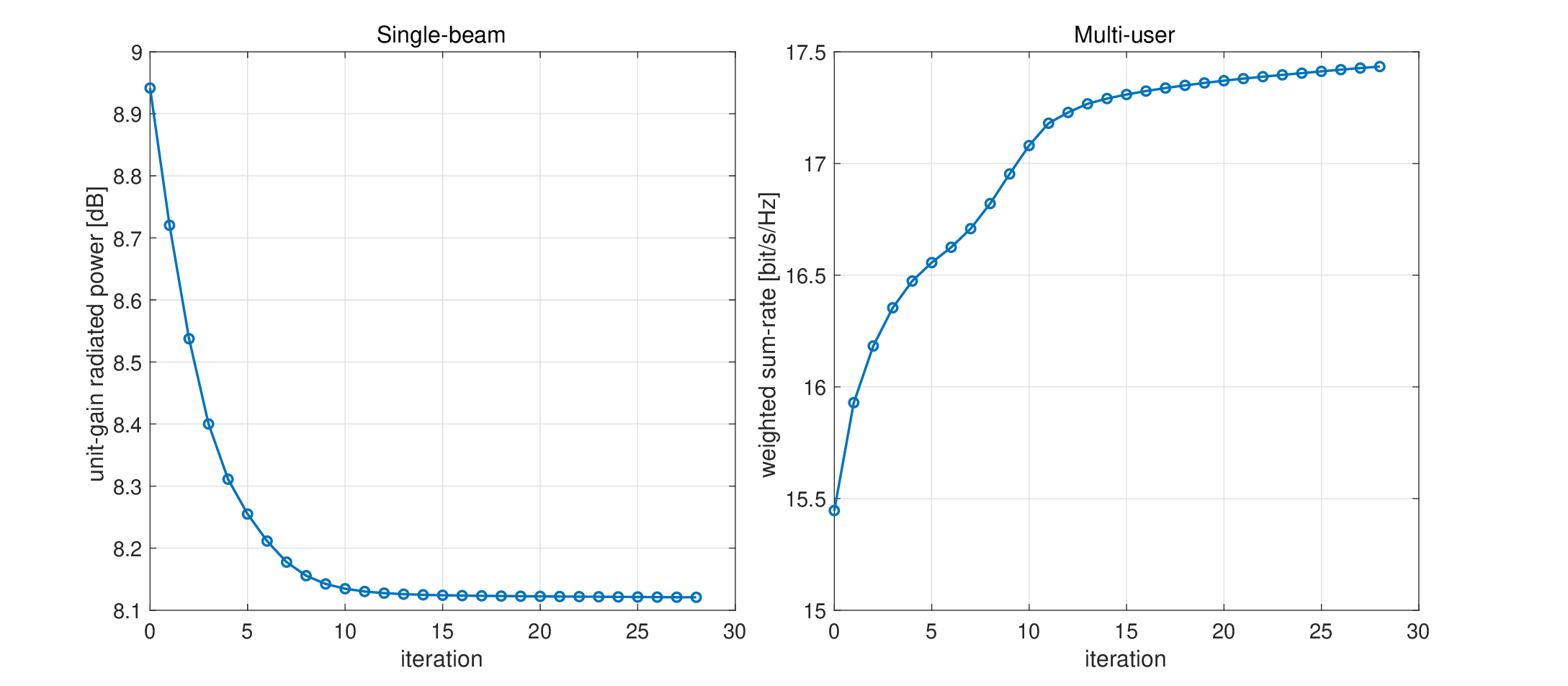}
	\caption{Convergence histories of the proposed optimization algorithm for the
		single-beam and multi-user cases. The left subfigure shows the unit-gain
		radiated power in the single-beam case, where a lower value is preferred.
		The right subfigure shows the weighted sum-rate in the multi-user case,
		where a higher value is preferred.}
	\label{fig:objective_history}
	\vspace{-4mm}
\end{figure}

Fig.~\ref{fig:beam_pattern}(a) shows the normalized azimuth beam patterns and
corresponding PSLL values. All schemes steer the main beam toward the desired
direction, yet their sidelobe behaviors differ markedly. The fixed-grid ideal
no-EM design exhibits the highest PSLL of $-6.0$ dB, indicating that neglecting
EM coupling degrades physical radiation performance when evaluated under the true
EM model. Incorporating EM coupling improves the PSLL to $-6.9$ dB, and
successively adding the current and voltage constraints further reduces it to
$-7.7$ dB and $-8.2$ dB, respectively—demonstrating that physical constraints
not only restrict the feasible set but also regularize the excitation currents
and suppress high-sidelobe patterns. The best random FAA attains $-7.8$ dB,
which is competitive but limited by the absence of directed geometry
optimization. The proposed optimized FAA achieves the lowest PSLL of $-9.0$ dB,
confirming that continuous port-position optimization provides a spatial degree
of freedom beyond current-only beamforming.

Fig.~\ref{fig:beam_pattern}(b) shows the 3-D normalized beam pattern of the
optimized FAA. A dominant main lobe is formed toward the target direction, while
the residual radiation is distributed into lower-level sidelobes, confirming
that the PSLL improvement is not a 2-D artifact but is consistent with the full
3-D radiation behavior. This highlights the key advantage of EM-aware FAS
co-design: jointly adapting port geometry and excitation currents enables
effective reshaping of both the main-lobe direction and the sidelobe structure
under realistic physical constraints.

\begin{figure}[t!]
	\centering
	\includegraphics[width=0.8\columnwidth]{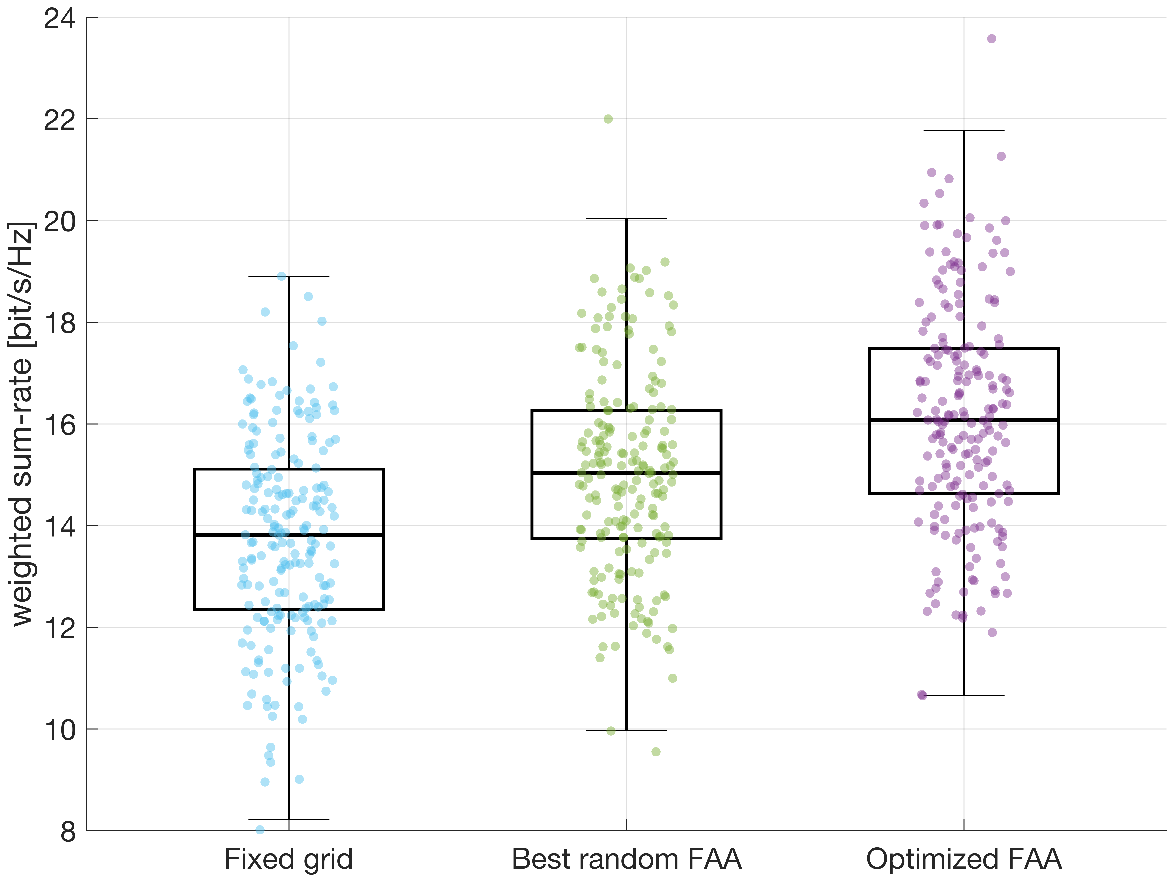}
	\caption{Weighted sum-rate statistics over 200 independent channel
		realizations. The three box plots compare the fixed-grid benchmark, the
		best random FAA, and the optimized FAA under the EM-aware multi-user
		precoding setting.}
	\label{fig:statistics}
\end{figure}

Fig.~\ref{fig:optimized_geometry} compares the optimized FAA geometries for the
two settings. In the single-beam case, several ports are driven toward the
aperture boundary, enlarging the effective aperture and sharpening the beam,
while the remaining interior ports provide degrees of freedom for sidelobe and
null control. In the multi-user case, the optimized geometry is more spatially
dispersed across the aperture, as expected, since the multi-user objective must
simultaneously enhance desired signal power, suppress inter-user interference,
account for EM coupling, and satisfy current and voltage feasibility. Unlike the
single-beam design, the multi-user geometry is accordingly adjusted to create a
more favorable channel structure for several angularly separated users.

Fig.~\ref{fig:objective_history} reports the convergence histories of both
optimizations. In the single-beam case, the unit-gain radiated power decreases
rapidly from $8.94$ dB to $8.12$ dB within the first 10 accepted outer
iterations and then flattens, indicating that most geometry gain is obtained
early while later iterations mainly refine port locations. In the multi-user
case, the weighted sum-rate increases monotonically from $15.45$ bit/s/Hz to
$17.45$ bit/s/Hz, with the main improvement within 12 iterations followed by
gradual saturation. The monotonic behavior in both cases confirms the stability
of the accepted-update strategy and validates the convergence of the proposed
alternating optimization.

Finally, Fig.~\ref{fig:statistics} presents the weighted sum-rate statistics
over 200 independent channel realizations. The fixed-grid benchmark yields the
lowest median of $13.8$ bit/s/Hz. Selecting the best feasible geometry from
random FAA realizations raises the median to $15.0$ bit/s/Hz, demonstrating
that geometry diversity alone provides a non-negligible gain over fixed
deployment. The proposed optimized FAA achieves the highest median of
$16.1$ bit/s/Hz, representing gains of $1.1$ bit/s/Hz over the best random FAA
and $2.3$ bit/s/Hz over the fixed-grid benchmark. The optimized-FAA distribution
is also visibly shifted upward, with the upper tail extending beyond
$20$ bit/s/Hz and the best observed realization reaching $23.6$ bit/s/Hz.
Although partial overlap exists in weaker channel realizations due to multipath
fading and user-interference variations, the upward shift in both the median and
upper quartile remains clear, confirming a statistically persistent sum-rate
advantage across diverse channel conditions.
\vspace{-4mm}
\section{Conclusion}
This paper presented an EM-aware current-domain framework for planar FAAs that
embeds mutual coupling, power, voltage, and geometry constraints into
communication design. The model was cast as two optimization problems:
single-beam superdirective beamforming and multi-user weighted sum-rate
maximization. The proposed alternating algorithms jointly optimize port currents
and positions while preserving physical feasibility. Simulation results
demonstrate that mutual coupling, when accurately modeled and properly
constrained, is not merely detrimental but can be leveraged to improve sidelobe
suppression and sum-rate performance.
\balance

\end{document}